\documentclass[aps, prd, twocolumn, amsmath, floats,floatfix, superscriptaddress, nofootinbib,
showpacs]{revtex4-1}

\usepackage{amssymb}
\usepackage{amsmath}
\usepackage{verbatim}
\usepackage{mathrsfs}
\usepackage{amsfonts}
\usepackage{latexsym}
\usepackage{epsfig}
\usepackage{color}
\usepackage{graphicx,subfigure}
\usepackage{units}
\usepackage{slashbox}
\usepackage{envmath}
\usepackage{natbib}

\usepackage[utf8]{inputenc}

\usepackage[unicode]{hyperref}
\hypersetup{
    unicode=true,          
    pdftitle={},
    pdfauthor={},     
    pdfsubject={},   
    pdfcreator={LaTeX},          
    pdfproducer={LaTeX},         
    pdfkeywords={}, 
    colorlinks=false,            
  }

\begin{document}


\definecolor{orange}{rgb}{0.9,0.45,0}

\newcommand{\re}{\mbox{Re}}
\newcommand{\im}{\mbox{Im}}
\newcommand{\tf}[1]{\textcolor{red}{#1}}
\newcommand{\jc}[1]{\textcolor{blue}{JC: #1}}
\newcommand{\nsg}[1]{\textcolor{cyan}{N: #1}}
\newcommand{\ch}[1]{\textcolor{green}{CH: #1}}

\def\CovDev{D}
\def\Res{{\mathcal R}}
\def\Gammaflat{\hat \Gamma}
\def\metricflat{\hat \gamma}
\def\Dflat{\hat {\mathcal D}}
\def\part_n{\partial_\perp}

\def\Lie{\mathcal{L}}
\def\A{\mathcal{X}}
\def\Aphi{\A_{\phi}}
\def\hAphi{\hat{\A}_{\phi}}
\def\E{\mathcal{E}}
\def\Ham{\mathcal{H}}
\def\M{\mathcal{M}}
\def\R{\mathcal{R}}
\def\p{\partial}

\def\hg{\hat{\gamma}}
\def\hA{\hat{A}}
\def\hD{\hat{D}}
\def\hE{\hat{E}}
\def\hR{\hat{R}}
\def\hcA{\hat{\mathcal{A}}}
\def\hDelt{\hat{\triangle}}

\renewcommand{\t}{\times}

\long\def\symbolfootnote[#1]#2{\begingroup%
\def\thefootnote{\fnsymbol{footnote}}\footnote[#1]{#2}\endgroup}


\title{Lensing and dynamics of ultra-compact bosonic stars} 	

\author{Pedro V. P. Cunha} 
\affiliation{Departamento de F\'{\i}sica da Universidade de Aveiro and 
Centre for Research and Development in Mathematics and Applications (CIDMA), 
Campus de Santiago, 
3810-183 Aveiro, Portugal}
\affiliation{CENTRA, Departamento de F\'{\i}sica, Instituto Superior T\'ecnico, Universidade de Lisboa, Avenida Rovisco Pais 1, 1049 Lisboa, Portugal}

\author{Jos\'e A. Font}
\affiliation{Departamento de
  Astronom\'{\i}a y Astrof\'{\i}sica, Universitat de Val\`encia,
  Dr. Moliner 50, 46100, Burjassot (Val\`encia), Spain}
\affiliation{Observatori Astron\`omic, Universitat de Val\`encia, C/ Catedr\'atico 
  Jos\'e Beltr\'an 2, 46980, Paterna (Val\`encia), Spain}

    \author{Carlos Herdeiro}
\affiliation{Departamento de F\'{\i}sica da Universidade de Aveiro and 
Centre for Research and Development in Mathematics and Applications (CIDMA), 
Campus de Santiago, 
3810-183 Aveiro, Portugal}

 \author{Eugen Radu}
\affiliation{Departamento de F\'{\i}sica da Universidade de Aveiro and 
Centre for Research and Development in Mathematics and Applications (CIDMA), 
Campus de Santiago, 
3810-183 Aveiro, Portugal}
  
   \author{Nicolas Sanchis-Gual}
\affiliation{Departamento de
  Astronom\'{\i}a y Astrof\'{\i}sica, Universitat de Val\`encia,
  Dr. Moliner 50, 46100, Burjassot (Val\`encia), Spain}
  
  \author{Miguel Zilh\~ao}
\affiliation{CENTRA, Departamento de F\'{\i}sica, Instituto Superior T\'ecnico, Universidade de Lisboa, Avenida Rovisco Pais 1, 1049 Lisboa, Portugal}
\affiliation{Departament de F\'{\i}sica Qu\`antica i Astrof\'{\i}sica \& Institut de Ci\`encies del Cosmos (ICC), Universitat de Barcelona, Mart\'{\i} i Franqu\`es 1, 08028 Barcelona, Spain}


\date{September 2017}


\begin{abstract} 
Spherically symmetric bosonic stars are one of the few examples of gravitating solitons that are known to form dynamically, via a classical process of (incomplete) gravitational collapse. As stationary solutions of the Einstein--Klein-Gordon or the Einstein--Proca theory, bosonic stars may also become sufficiently compact to develop light rings and hence mimic, in principle, gravitational-wave observational signatures of black holes (BHs). In this paper, we discuss how these horizonless \textit{ultra-compact objects} (UCOs) are actually distinct from BHs, both phenomenologically and dynamically. In the electromagnetic channel, the light ring associated phenomenology  reveals remarkable lensing patterns, quite distinct from a standard BH shadow, with an infinite number of Einstein rings accumulating in the vicinity of the light ring, \textit{both inside and outside} the latter. The strong lensing region, moreover, can be considerably smaller than the shadow of a BH with a comparable mass. Dynamically, we investigate the fate of such UCOs under perturbations, via fully non-linear numerical simulations and observe that, in all cases,  they decay into a Schwarzschild BH within a time scale of $\mathcal{O}(M)$, where $M$ is the mass of the bosonic star. Both these studies reinforce how difficult it is for horizonless UCOs to mimic BH phenomenology and dynamics, in all its aspects. 
\end{abstract}




\maketitle

\vspace{0.8cm}

\section{Introduction}

The true nature of astrophysical black hole (BH) candidates has been a central question in relativistic astrophysics for decades. The observational elusiveness of their defining property -- the existence of an event horizon --, allows the possibility that they may, in reality, be some sort of exotic \textit{horizonless} compact objects, whose phenomenology is sufficiently similar to that of BHs, so that current observations are unable to distinguish these two types of objects.

In this context,  the recently opened gravitational-wave window to the Cosmos~\cite{Abbott:2016blz,Abbott:2016nmj,Abbott:2017vtc}, offers a particularly well suited channel to probe the nature of compact objects. Yet, it has been recently emphasised that observational degeneracy may still remain in this channel~\cite{Cardoso:2016rao}.  The correspondence between a BH's natural oscillation frequencies (so called quasi-normal modes~\cite{Berti:2009kk}) and light ring (LR) vibrations~\cite{1972ApJ...172L..95G,Cardoso:2008bp,Khanna:2016yow}, implies that compact objects with a LR -- henceforth \textit{ultra-compact objects} (UCOs) -- but with no event horizon can mimic the initial part of the ringdown gravitational-wave signal of perturbed BHs. Later parts of the ringdown signal may have signatures of the true nature of the object (through the so called \textit{echos}~\cite{Cardoso:2016oxy,Cardoso:2017njb}), but the corresponding lower signal to noise ratio challenges clean detections of this part of the signal, at least in the near future -- see~\cite{Abedi:2016hgu,Ashton:2016xff,Price:2017cjr,Nakano:2017fvh,Mark:2017dnq} for recent discussions.

Is there, consequently, a real risk of observationally mistaking UCOs by BHs and vice-versa, with current and near future gravitational-wave measurements? To address this important question, one should start by revisiting the theoretical foundations of concrete UCOs models. Even though many variants of horizonless UCOs have been proposed in the literature, either as stationary solutions of well-defined models or as more speculative possibilities (see $e.g.$~\cite{Visser:1995cc,Schunck:2003kk,Mathur:2005zp,Mazur:2004fk,Gimon:2007ur,Brito:2015pxa}), they generically suffer from the absence of a plausible formation scenario. An exception, in this respect, are (scalar) boson stars, which, in spherical symmetry, have been shown to form from a process of gravitational collapse, due to an efficient cooling mechanism~\cite{Seidel:1993zk}. Moreover, boson stars are known to become UCOs, in parts of their domain of existence~\cite{Cunha:2015yba}. 

In this paper, we shall take spherically symmetric scalar boson stars, as well as their vector cousins, dubbed \textit{Proca stars}~\cite{Brito:2015pxa}, collectively referred to as \textit{bosonic stars}, as a reference example of horizonless UCOs, and simultaneously as a proof of concept that BH mimickers are dynamically possible through known physics.  We then aim at assessing their quality as BH mimickers by performing the following two inquires: $1)$ when in the UCO regime, does \textit{all their LR associated phenomenology} mimic that of a Schwarzschild BH? $2)$ if perturbed, do they really oscillate as a Schwarzschild BH?

Our study reveals that bosonic stars, both the scalar and the vector ones, fail to pass either of these tests. Firstly, the same LR that allows them to, in principle, vibrate as BHs do, gives rise to a quite distinct pattern of light lensing from standard BH shadows. In a sense, the LR associated electromagnetic channel phenomenology raises the degeneracy of the gravitational channel phenomenology. Secondly, and more importantly, bosonic stars only become UCOs in a regime wherein they are also perturbatively unstable. Thus, the same perturbations that could make them vibrate as a BH will actually induce their gravitational collapse into one. By performing fully non-linear simulations we show that this is a fast process, and a horizon forms within a few light-crossing times. All together, these results emphasise the difficulty, at least in spherical symmetry, in constructing a reasonable dynamical model of horizonless UCOs whose phenomenology can mimic that of a BH, in all its aspects. 

This paper is organised as follows. In section~\ref{sec2} we review the main physical properties of the ultra-compact bosonic stars we shall be analysing. In section~\ref{sec3} we shall analyse their lensing, and compare it with that of a Schwarzschild BH. In section~\ref{sec4} we consider their behaviour under perturbations, following, fully non-linearly, their evolution and collapse into a BH. In section~\ref{sec5} we present our final remarks. One appendix addresses the 3+1 formalism used in section~\ref{sec42}.

\section{Ultra-compact  bosonic stars}
\label{sec2}
The ultra-compact bosonic stars we shall be considering in this paper are solutions of Einstein's gravity minimally coupled with a spin-$s$ field, with $s=0,1$. The scalar case was first discussed in~\cite{Kaup:1968zz,Ruffini:1969qy} and it is reviewed in~\cite{Schunck:2003kk}. The vector case was first discussed in~\cite{Brito:2015pxa}. The models are summarised by the action (we use units with $c=1=\hbar$ and $4\pi G=1$)
\begin{eqnarray}
\label{action}
\mathcal{S}=\int d^4 x \sqrt{-g} 
\left [
\frac{R}{4} 
+
\mathcal{L}_{(s)}
\right] \ ,
\end{eqnarray}
where the two corresponding matter Lagrangians are:
\begin{eqnarray}
\label{lagrangians}
&&\mathcal{L}_{(0)}= - g^{\alpha \beta}\bar{\Phi}_{, \, \alpha} \Phi_{, \, \beta} - \mu^2 \bar{\Phi}\Phi \ , \\ 
&&\mathcal{L}_{(1)}= -\frac{1}{4}\bar{\mathcal{F}}_{\alpha\beta}{\mathcal{F}}^{\alpha\beta}
-\frac{1}{2}\mu^2\bar{\mathcal{A}}_\alpha {\mathcal{A}}^\alpha \ .
\end{eqnarray}
Here, $\Phi$ is a complex scalar field and $\mathcal{A}$ is a complex 4-potential, with field strength $\mathcal{F} =d\mathcal{A}$. The overbar denotes complex conjugate and $\mu$ is the field's mass.  In this paper, the conventions for scalars and vectors are those in~\cite{Herdeiro:2017fhv} (see also~\cite{Herdeiro:2015gia,Brito:2015pxa,Herdeiro:2016tmi}).

We shall be interested in spherically symmetric solutions. They are obtained using the line element
\begin{equation}
\label{ansatz1}
 ds^2=-N(r)\sigma^2(r) dt^2+\frac{dr^2}{N(r)}+r^2 (d\theta^2+\sin^2\theta d\varphi^2) \ ,
\end{equation} 
where $N(r)\equiv 1-{2m(r)}/{r}$, $m(r),\sigma(r)$ are radial functions and $r,\theta,\varphi$ correspond  to 
Schwarzschild-type coordinates. In particular the radial coordinate $r$ is the geometrically meaningful \textit{areal radius}, meaning that the proper area of a 2-sphere ($r,t=$constant) is $4\pi r^2$.  The matter fields ansatz is given in terms of another real function $\phi(r)$ [two 
real functions $(V(r),H_1(r))$], for the scalar [vector] case:
\begin{equation}
\label{ansatzfield}
\Phi=\phi(r)e^{-iwt} \ , \ \ 
\mathcal{A}=\left[ iV(r) dt+
\frac{H_1(r)}{r}dr   
\right] e^{-iw t} \ ,
  \ 
\end{equation}
where $w>0$ is the frequency of the field.
The Einstein-matter equations are solved, numerically, with appropriate boundary conditions. The explicit form of these equations and boundary conditions, together with  some examples of profiles of the matter and metric functions can be found in~\cite{Herdeiro:2017fhv} (see also~\cite{Brito:2015pxa} for the Proca case). 

\begin{figure*}[tbhp]
\centering
\includegraphics[height=2.4in]{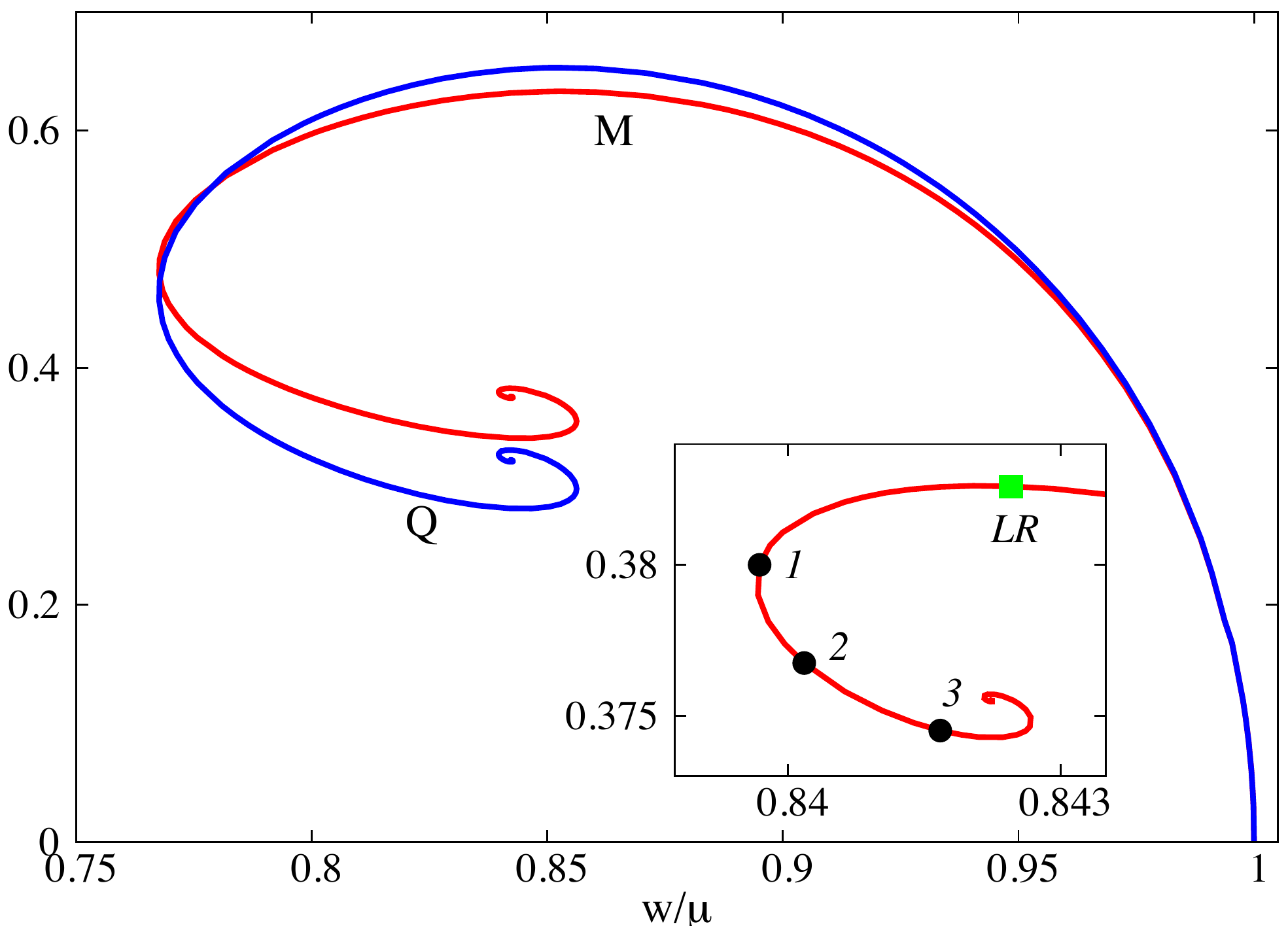}  \ \ \ 
\includegraphics[height=2.4in]{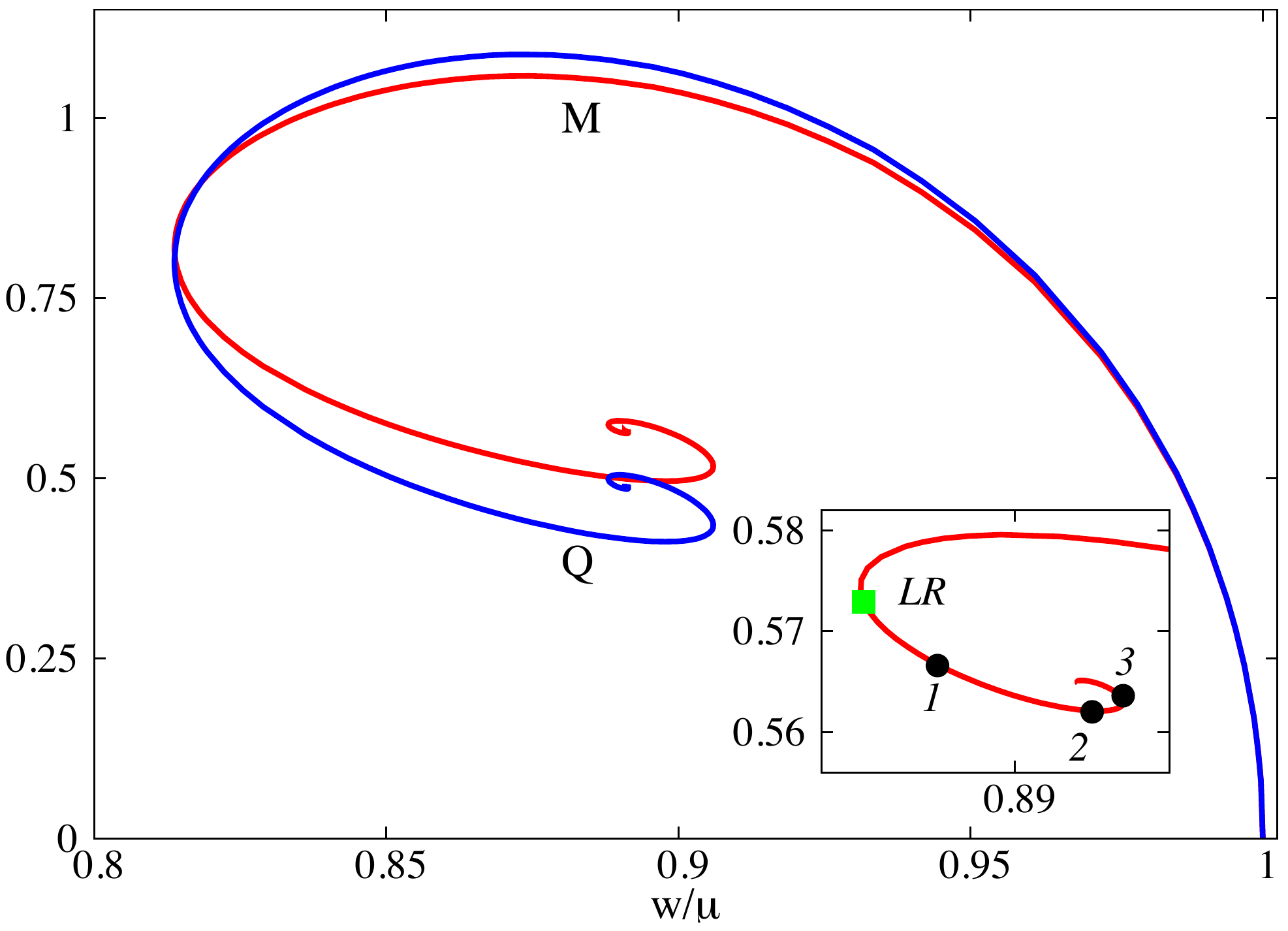}   \vspace{0.4cm} \\
\includegraphics[height=2.4in]{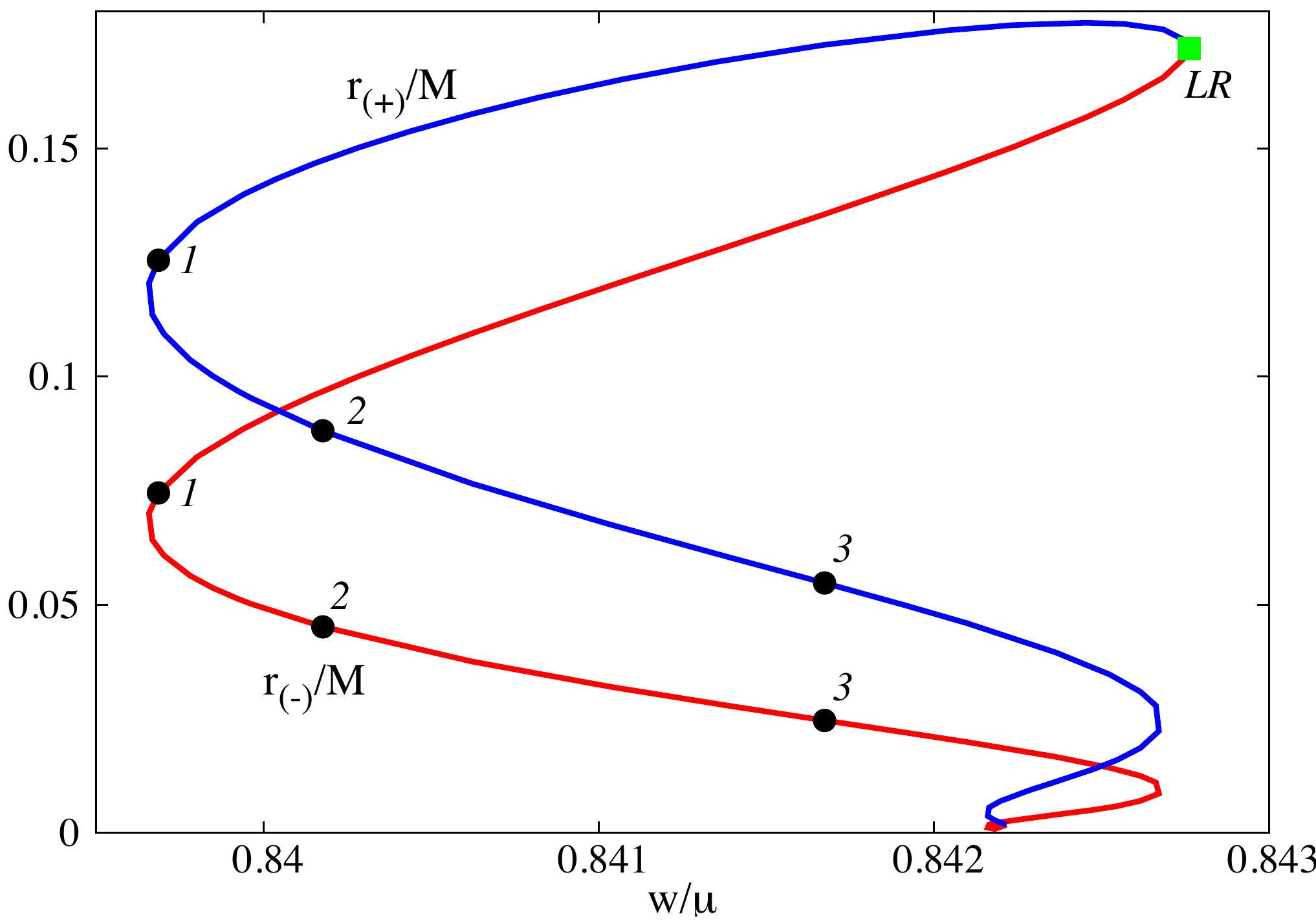}  \ \ \ 
\includegraphics[height=2.4in]{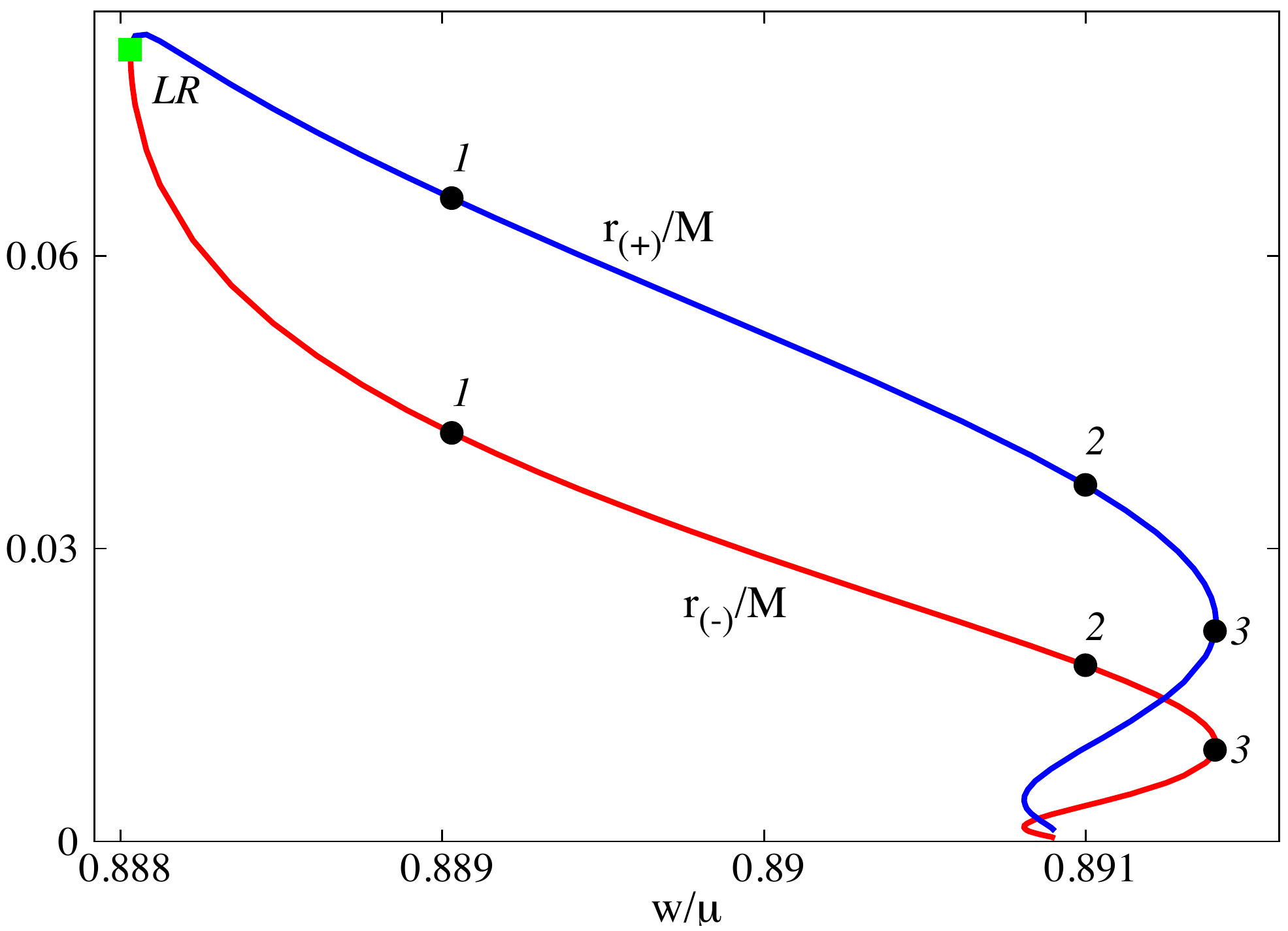} \vspace{0.4cm} \\
\includegraphics[height=2.4in]{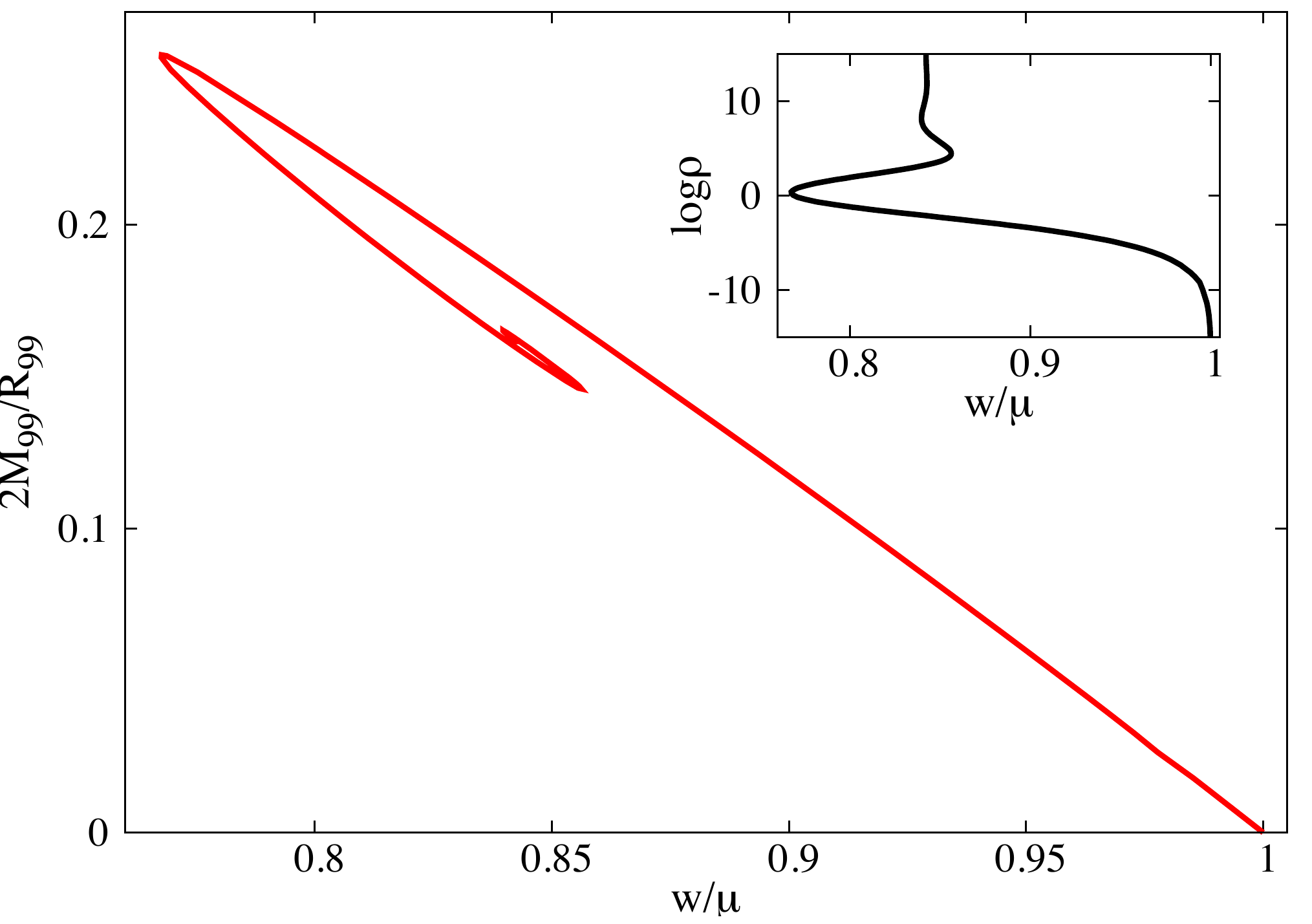}  \ \ \ 
\includegraphics[height=2.4in]{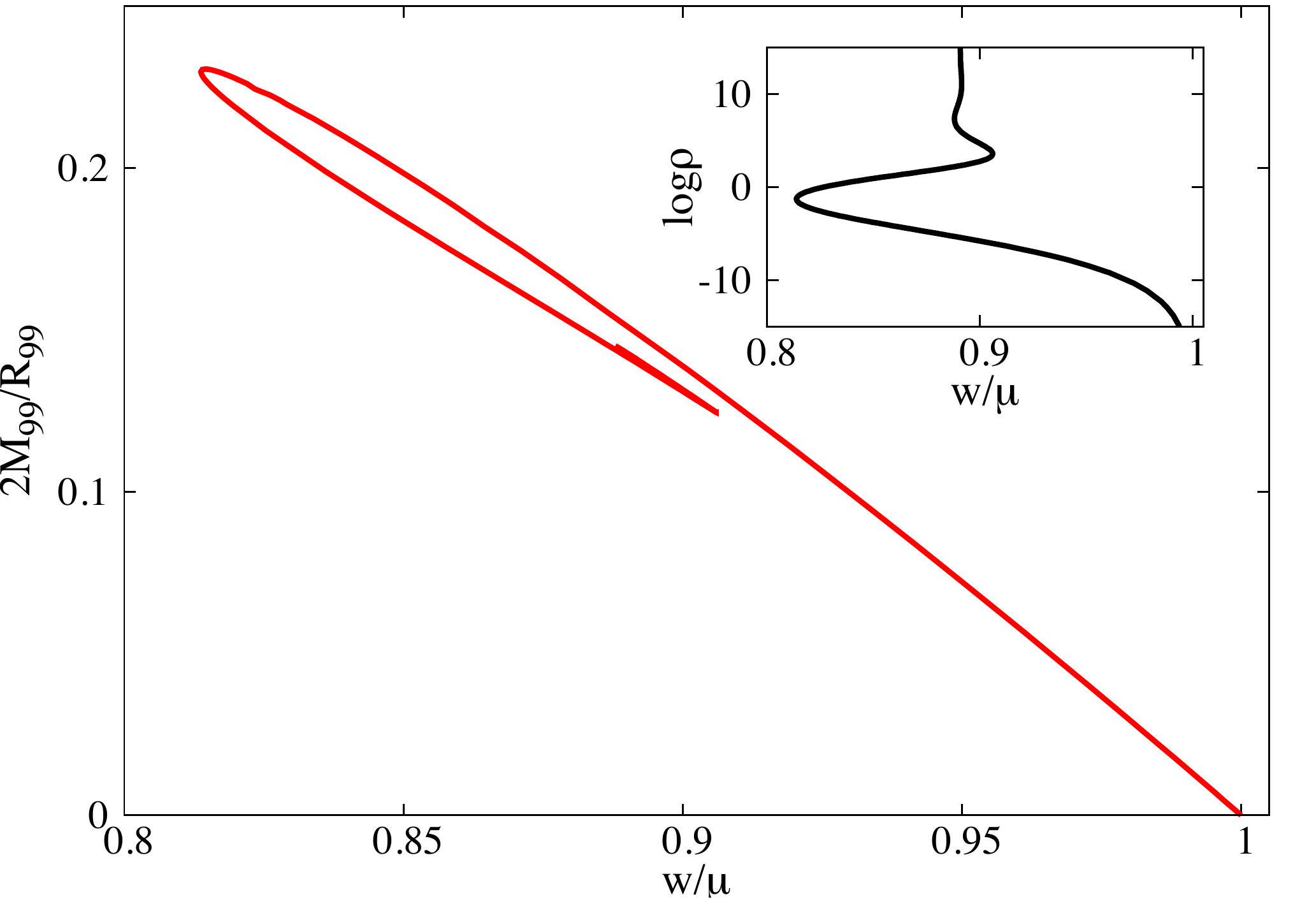} 

\caption{(Top panels) Domain of existence of scalar boson stars (left) and Proca stars (right) in an ADM mass (M)/Noether charge (Q) $vs.$ field frequency, $w/\mu$,
diagram. The green square marks the first solution with a LR. The three highlighted points correspond to the configurations we have analysed in detail, in each case. (Middle panels) Areal radius of the inner $r_-$ and outer $r_+$ LRs, normalised to the ADM mass, as a function of $w$, in the region where LRs appear. (Bottom panels) Compactness of the scalar boson stars (left) and Proca stars (right), as measured by $M_{99}/2R_{99}$ (see main text). The inset shows the (log of the) central density. Observe that $\rho$ can get extremely large in the central region, although the solutions will not get more compact, as measured by $M_{99}/2R_{99}$.}
\label{fig1}
\end{figure*}

In Fig.~\ref{fig1} we exhibit various properties of the scalar (left columns) and vector (right columns) bosonic stars which are relevant for our study. The top panels show the domain of existence of the solutions in an ADM mass, $M$, $vs.$ a bosonic field frequency, $w$, diagram. Regardless of the spin, the solutions form a characteristic spiralling curve, starting from the Newtonian regime (as $w\rightarrow \mu$) wherein the bosonic stars tend to become dilute and weakly relativistic. Following the spiral from this Newtonian limit, the ADM mass reaches a maximum at some frequency. These maximal mass and corresponding frequency are, in units with $\mu=1$, $(M_{\rm max},w[M_{\rm max}])=(0.633,0.853)$ for the scalar case and $(M_{\rm max},w[M_{\rm max}])=(1.058,0.875)$ for the vector case. Perturbation theory computations for both the scalar~\cite{Gleiser:1988ih,Lee:1988av} and vector cases~\cite{Brito:2015pxa} have shown that at this point in the spiral an unstable mode develops. More relativistic solutions become perturbatively unstable with different possible fates~\cite{Seidel:1990jh,Sanchis-Gual:2017bhw}. Further following the spiral, one finds several backbendings, each defining the end of a \textit{branch}. As it can be seen in the inset of the top panels, the solution at which a LR is first seen (marked by a green square -- see~\cite{Cunha:2015yba,Cunha:2016bjh,Cunha:2017eoe} for quantitative details) occurs in the third (fourth) branch for the scalar (vector) case, corresponding to $(M_{\rm LR},w[M_{\rm LR}])=(0.8424,0.383)$ for the scalar case and $(M_{\rm LR},w[M_{\rm LR}])=(0.8880,0.573)$ for the vector case. These are highly relativistic solutions, with redshift factors approaching those of an event horizon towards the centre of the solutions. In each case, we have highlighted three solutions, denoted 1-3, in the insets of the top panels of Fig.~\ref{fig1}, corresponding to the solutions we shall analyse below in more detail. These are ultra-compact solutions and their physical properties are summarised in Table~\ref{tab:mod1}. The top panels of Fig.~\ref{fig1} also show the Noether charge of the solutions, $Q$ (see $e.g.$~\cite{Herdeiro:2017fhv} for quantitative expressions), corresponding to a locally conserved charge associated with the global $U(1)$ symmetry of each family of solutions. The ratio $Q/M$, in units with $\mu=1$, provides a criterion for stability:  $Q/M<1$ implies excess energy and hence instability against fission into unbound bosonic particles. The point at which, in both cases, solutions have excess (rather than binding) energy occurs close to the minimum frequency, and thus already in the region of perturbative instability.

The middle panels in~Fig.~\ref{fig1} exhibit the value of the areal radius of each LR, in units of the ADM mass, and its variation along the ultra-compact bosonic star solutions. When the LR first appears in the spiral representing the family of bosonic star solutions it is actually \textit{degenerate}. This solution marks the beginning of the ultra-compact bosonic stars. Deeper into the centre of the spiral, the bosonic stars have two LRs; in fact, generically, smooth ultra-compact objects have an even number of LRs~\cite{Cunha:2017qtt}.  The outermost one (with radial coordinate $r_{(+)}$, blue line) always corresponds to an unstable photon orbit; the innermost (with radial coordinate $r_{(-)}$, red line) always corresponds to a stable orbit~\cite{Cunha:2017qtt}. As the figure shows, the two radial coordinates start to bifurcate from the first ultra-compact solution, but then converge again, towards the centre of the spiral. Interestingly, the areal radius of the LRs is much smaller than that of a Schwarzschild BH, for which $r/M=3$. This is associated with the fact these solutions are not constant-density stars, having a much denser central region (inset of bottom panels of~Fig.~\ref{fig1}). 
The three chosen solutions are also highlighted in these plots, and the corresponding LRs areal radii are given in Table I.

\begin{table}[h!]
\caption{Ultra-compact bosonic star models.}
\label{tab:mod1}
\begin{ruledtabular}
\begin{tabular}{cccccc}
Model&$w/\mu$&$\mu M_{\rm ADM}$&$\mu^2Q$&$r_{(-)}/M$&$r_{(+)}/M$\\
\hline
BS1&0.8397&0.3800&0.3274 &0.074&0.126\\
BS2&0.8402&0.3767&0.3235 &0.045&0.088\\
BS3&0.8417&0.3745&0.3209 &0.024&0.053\\
PS1&0.8890&0.5666&0.4899 &0.042&0.065\\
PS2&0.8911&0.5621 &0.4849 &0.016&0.034\\
PS3&0.8914&0.5636&0.4866 &0.007&0.018\\
\end{tabular}
\end{ruledtabular}
\end{table}

The bottom panel in~Fig.~\ref{fig1} show a measure of the compactness of the bosonic stars. Since these stars have no hard surface, several measures of compactness are possible. In view of their exponential fall-off of the matter density, following, $e.g.$~\cite{AmaroSeoane:2010qx,Herdeiro:2015gia}, we have defined compactness as the ratio of the Schwarzschild radius for 99\% of the mass, denoted $2M_{99}$, to the areal radius that contains such mass, $R_{99}$. This quantity would be unity for a Schwarzschild BH. Here we see that the compactness increases from the Newtonian limit until the first back bending, but it decreases along the second branch. Then it increases along the third branch. Such compactness is not a monotonic function along the spiral and indeed the ultra-compact solutions -- in the sense of possessing a LR -- are not the most compact ones, according to this definition. On the other hand, the central value of the energy density (see $e.g.$~\cite{Herdeiro:2017fhv} for quantitative expressions) is indeed a monotonically increasing function along the spiral, as shown in the inset of these plots. This behaviour, together with the location of the LRs, show that for non-constant density stars, like these bosonic stars, a global measure of compactness, such as $2M_{99}/R_{99}$, may be misleading, as the star may have a considerably denser central region, which is ultra-compact, whereas the star as a whole is not. 


\section{Lensing}
\label{sec3}

LRs are bound planar photon orbits (see~\cite{Cunha:2017eoe} for a general discussion of bound photon orbits). Their existence around a compact object implies strong lensing effects. For the Schwarzschild BH, the LR occurs at an areal radius $r=3M$ and it is an \textit{unstable} photon orbit. Thus, scattering photons with an impact parameter  ($\eta=L/E$, where $E,L$ are the photon's energy and angular momentum, respectively) larger than (in modulus) that of the LR, $\eta_{LR}$, return to spatial infinity; but, when $\eta$ is close to $\eta_{LR}$, $|\eta|\gtrsim |\eta_{LR}|$, the scattering angle can be arbitrarily large, in the sense that the photon may circumnavigate the BH an arbitrary number of times before bouncing back to infinity. If $|\eta|< |\eta_{LR}|$ , on the other hand, the photon will end up falling into the BH. Thus the LR, defines an absorption cross section for light, the \textit{BH shadow}~\cite{1973blho.conf..215B,Falcke:1999pj}. This is a timely observable, due to ongoing attempts to measure the BH shadow of two supermassive BHs, by the Event Horizon Telescope~\cite{2009astro2010S..68D,Lu:2014zja}.

\begin{figure*}[tbhp]
\centering
\includegraphics[height=3.4in]{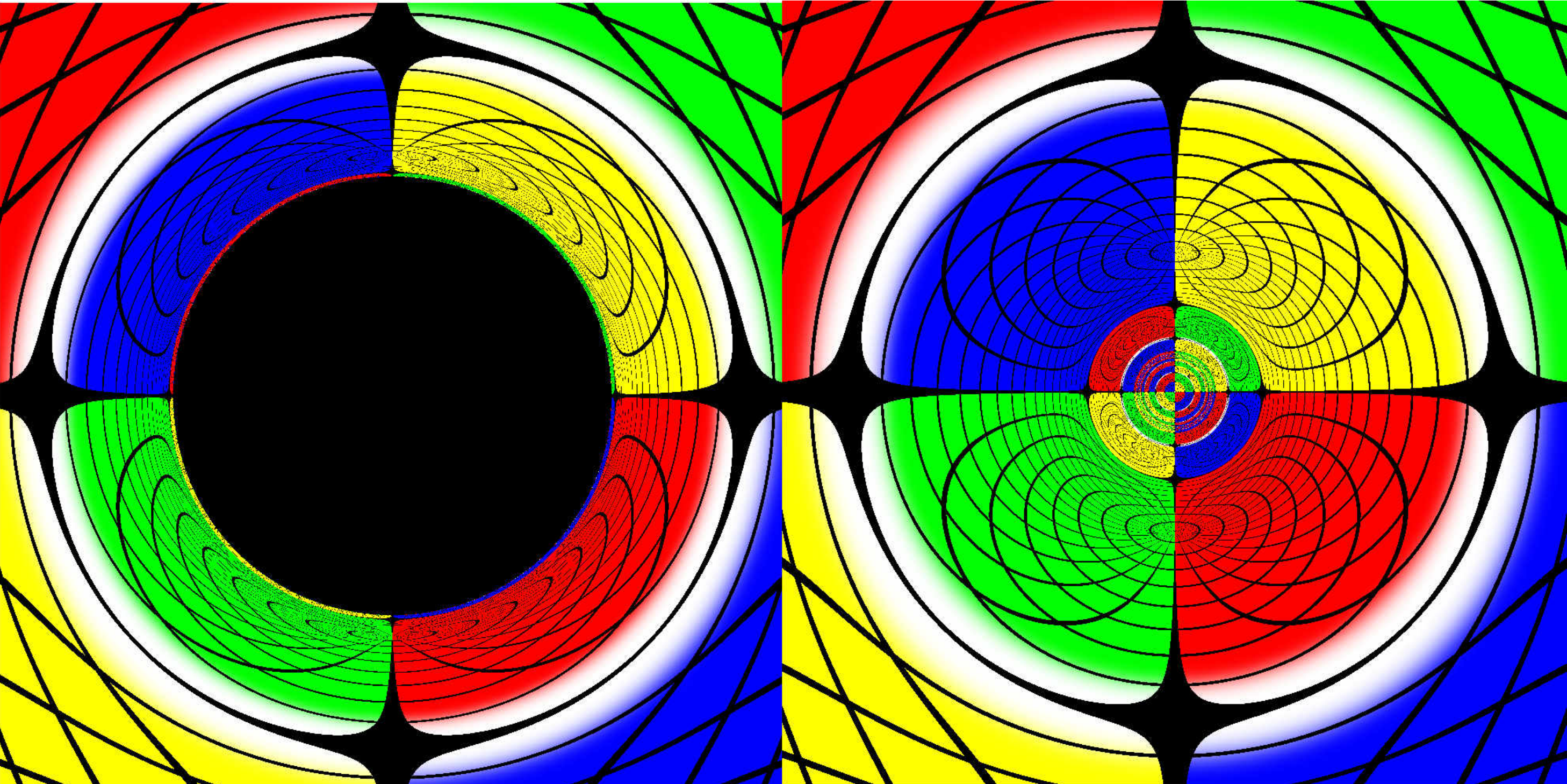}    
\caption{Lensing and shadow of a Schwarzschild BH (left panel) and a comparable bosonic star (right panel, model PS2), in similar observation conditions, for which the observer is set at $r_{\rm obs}=15M$. The Einstein ring has a similar dimension (white lensed region), but the strongly lensed region -- shadow and near its edge for the BH $vs.$ central rings for the star -- is much smaller for the star. 
}
\label{fig2}
\end{figure*}

\begin{figure*}[tbhp]
\centering
\includegraphics[height=3.4in]{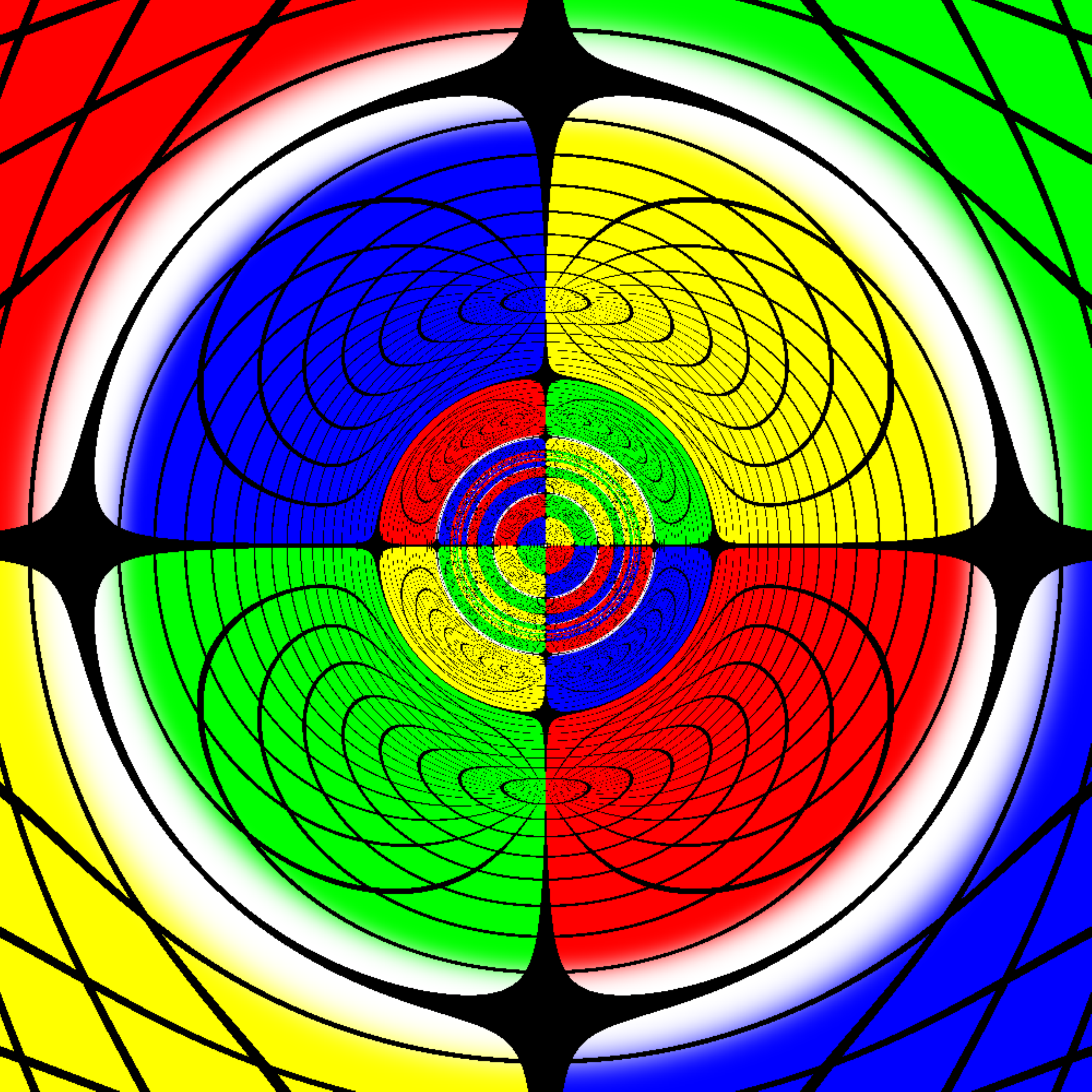}  
\includegraphics[height=3.4in]{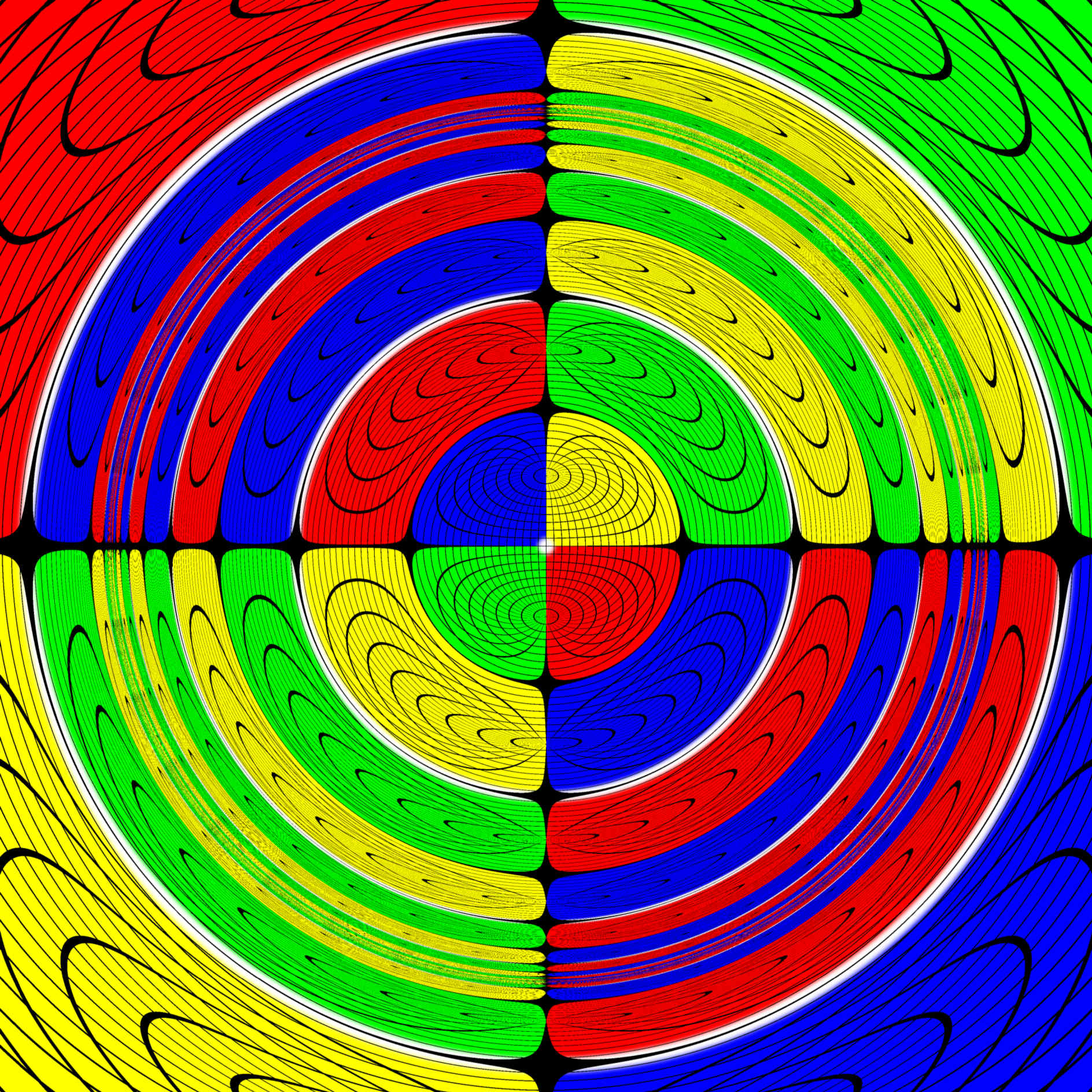}  
\caption{Lensing by the boson star model BS1 (left panel) and a zoom around the strong lensing region (right panel).
}
\label{fig3}
\end{figure*}

 In Fig.~\ref{fig2} (left panel) the BH shadow and lensing due to a Schwarzschild BH is shown. The setup is the one introduced in~\cite{Bohn:2014xxa} and used by some of us in~\cite{Cunha:2015yba,Cunha:2016bjh,Cunha:2017eoe}, wherein the  numerical ray-tracing method is also described. In a nutshell, light emanates from a far away celestial sphere that is divided into four quadrants, each painted with one colour (yellow, blue, green red). Black constant latitude and longitude lines are also drawn in the light-emitting celestial sphere. The observer is placed off-centre, within the celestial sphere at some areal radius $r_{\rm obs}$. Directly in front of the observer, there is a point in the celestial sphere where the four quadrants meet, which is painted in white and blurred. The Schwarzschild BH is placed at the centre of the celestial sphere. 
 
 The left panel of Fig.~\ref{fig2} has a few distinctive features. The white circle is the lensing, due to the BH, of the celestial sphere's white dot, which would be right in front of the observer if the BH would not be in the line of sight. It is an \textit{Einstein ring}~\cite{Einstein:1956zz} -- see~\cite{1997Sci...275..184R} for an historical account of the prediction of multiple images of a source due to gravitational lensing. The black central disk is the BH shadow, whose edge corresponds to photons that skim the LR. In between this edge and the Einstein ring there are infinitely many copies of the celestial sphere, that accumulate in the neighbourhood of the shadow's edge, in a self-similar structure~\cite{Bohn:2014xxa}. In the image only two of these copies are clearly visible. 

The right panel of Fig.~\ref{fig2} shows the lensing pattern due to a bosonic star, model PS2, under similar observation conditions, $i.e.$ and observer placed at the same $r_{\rm obs}$ and the BH replaced by the star at the centre of the celestial sphere. Since the $g_{tt}$ component of the metric is very close to zero within the star region, the numerical integration of the null geodesics is quite demanding. This issue is tamed by performing a conformal transformation to a spacetime with less extreme redshift factor,  since such transformation leaves invariant null geodesic paths. This is an efficient procedure. We have checked different conformal transformations lead to the same image, validating the method.

Comparing the left and right panels of Fig.~\ref{fig2}, leads to two main conclusions. Firstly, the Einstein ring has a similar dimension. Since there is only one scale for either solution -- the total mass -- similar observation conditions imply the lensing is due to objects with the same total mass. This explains the same overall (weak) light bending that originates the Einstein ring. Secondly, the strong lensing region, which is due to photons with $\eta\sim \eta_{LR}$, is smaller for the star. This is a consequence of the smaller LRs, $cf.$ the previous section: for ultra-compact bosonic stars they occur at an areal radius $\ll 3M$.

The lensing for the six selected models of ultra-compact bosonic stars ($cf.$ Table I) is qualitatively similar. In Fig.~\ref{fig3} (left panel), we exhibit the one for model BS1, under similar observation conditions $r_{\rm obs}=15M$ as the one for PS2 shown in the right panel of Fig.~\ref{fig2}. As expected the Einstein ring has a similar scale, but the strong lensing region is smaller for the Proca star, which is, qualitatively, in agreement with its smaller (outer) LR. It is important to emphasise, however, that the angular size in the image is determined by the LR's impact parameter, and not by its areal radius~\cite{Cunha:2016wzk}. 

\begin{figure}[h!]
\centering
\includegraphics[height=2.5in]{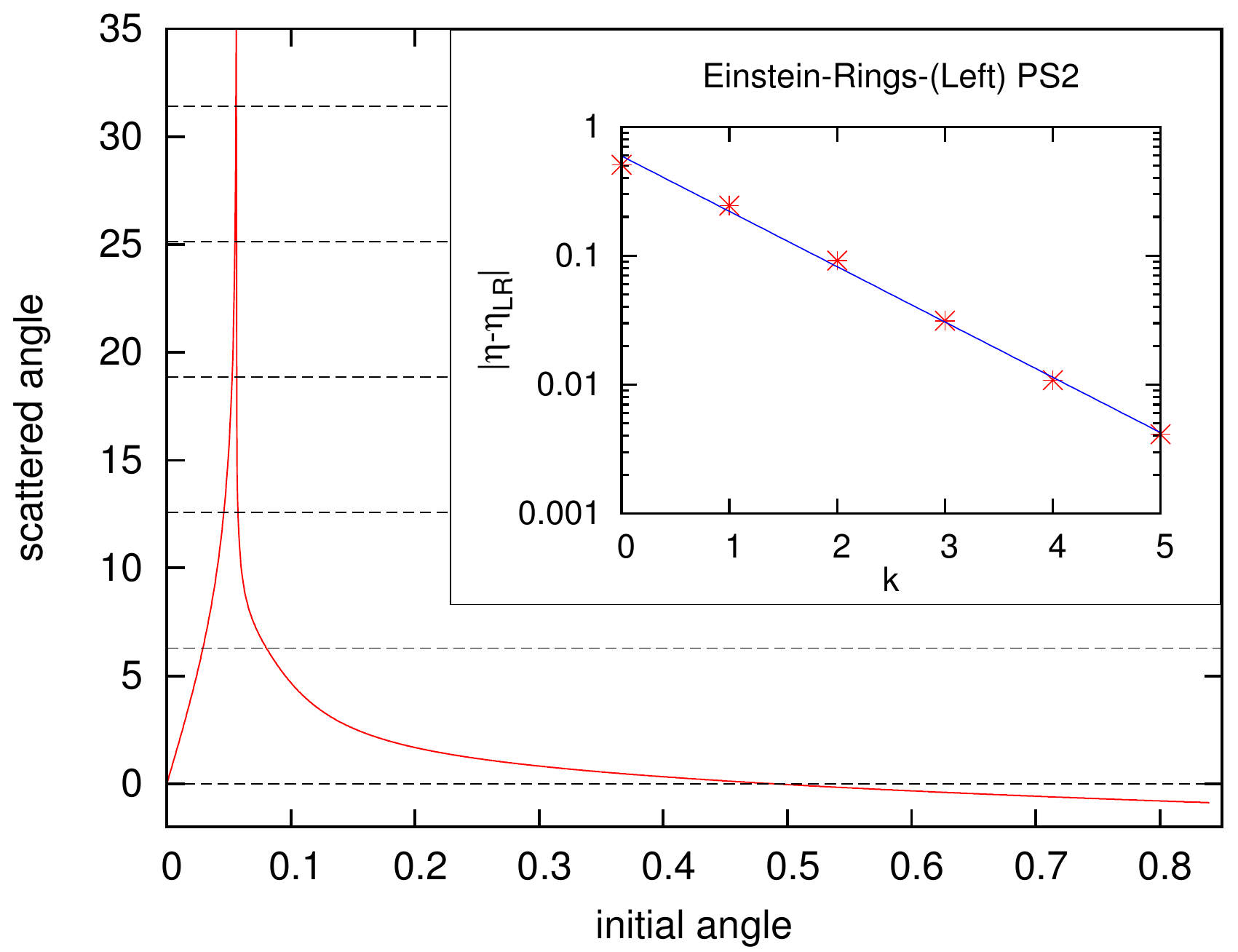}  
\caption{(Main panel) Scattering angle $vs.$ initial angle for the scattered photons in model PS2. (Inset) Value of $|\eta_{\rm ER}^{(k)} - \eta_{\rm LR}|$  given by eq.~\eqref{ERap}  (blue solid line) $vs.$ numerical values (star-like points) for the first five Einstein rings in model PS2.}
\label{fig4}
\end{figure}

The right panel of Fig.~\ref{fig3} shows a zoom of the left panel, around its central region. Circles, which are Einstein rings, are the lensing images of either the celestial sphere point in front of the observer (white circles) or the one behind the observer (black circles). These two types of circles alternate and appear to accumulate at a given angular radius. This can be confirmed in Fig.~\ref{fig4} (main panel), which displays the initial angle (which one can regard as the radial coordinate in the lensing images) $vs.$ the scattering angle, $i.e.$ the final angle in the celestial sphere. The scattering angle is here taken to be zero at the white dot of the celestial sphere (directly on the observer's line of sight, if the geometry were flat). Hence, multiples of $2\pi$ signal the formation of a white circle in the image, which can be seen by the horizontal dashed lines in Fig.~\ref{fig4}. The peak on the plot is the finger print of the unstable LR. Had this been a BH, instead of a bosonic star, the left part of the peak would not exist, as it would correspond to the shadow. The region in between each consecutive black and a white circles in Fig.~\ref{fig3}, contains a copy-image of the full celestial sphere. As familiar from particle physics/quantum mechanics, the outermost (unstable) LR, which is a bound state, appears as a pole in the scattering amplitude -- Fig.~\ref{fig4}.

The scattering angle divergence near the LR is logarithmic. This allows us to write the impact parameter of the Einstein ring of order $k$, corresponding to a scattering angle of $2\pi k$ as
\begin{equation}
\eta_{\rm ER}^{(k)} = \eta_{\rm LR}  +  b e^{-2\pi k/a}\ ,
\label{ERap}
\end{equation}
where $a,b$ are constants, the value of which depends on the LR being
approached with values of $\eta$ above or below $\eta_{\rm LR}$. Fig.~\ref{fig4} (inset) shows this relation is a good approximation to the numerical values, even for the lowest order Einstein rings.

\begin{figure}[h!]
\centering
\includegraphics[height=3.4in]{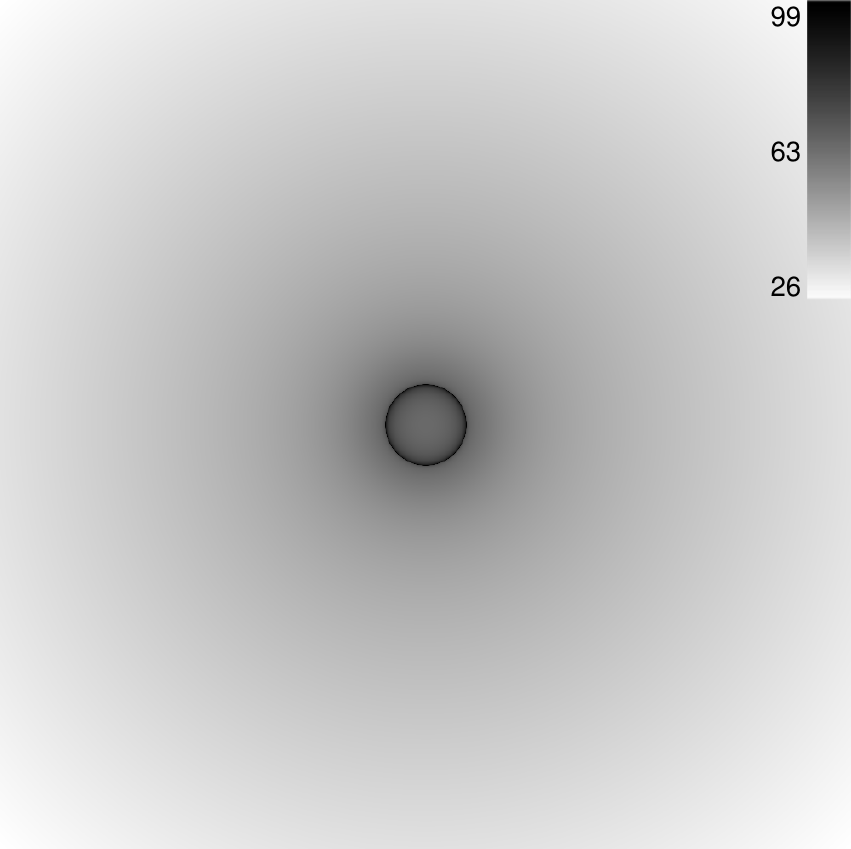}  
\caption{Time delay heat map for model PS2.}
\label{fig5}
\end{figure}

Whereas the LR is not emphasised in the plots in Figs.~\ref{fig2} and~\ref{fig3}, it stands out if instead we plot the \textit{time delay function}. This function is defined as the variation of the coordinate time $t$,  in units of $M$, required for the photon geodesic emanating from a particular pixel to reach a corresponding point on the celestial sphere~\cite{Cunha:2016bjh}.  This is a good diagnosis of the LR since photon trajectories that skim the LR take much longer to return to spatial infinity.  In Fig.~\ref{fig5} the time delay  for model PS2 is portrayed as a heat map with the corresponding scale on the right of the image indicating the variation of the coordinate time (in units of $M$) for each photon to travel from the camera to the celestial sphere.  The LR clearly stands out (compare with Fig.~\ref{fig2}, right panel). 

Figure~\ref{fig5} shows that UCOs like bosonic stars -- made of dark matter that only affect light
through the spacetime geometry -- have a ring-like darker region, rather than a disk-like shadow. Of course, it is possible that in a more realistic astrophysical environment, with an accretion disk light source, the whole central region becomes an effective shadow, $cf.$~\cite{Vincent:2015xta}. Likely, this depends on the accretion modelling and, in any case, this effective shadow will be considerably smaller than that of a comparable Schwarzschild BH. 

Let us close this section by remarking that the innermost (stable) LR plays no significant role in the photon scattering problem we have just analysed, but it potentially impacts in the spacetime stability~\cite{Keir:2014oka,Cardoso:2014sna,Cunha:2017qtt}. 

\section{Non-linear evolutions} 
\label{sec4}

We now turn to the dynamical evolution of ultra-compact bosonic stars. As already mentioned, ultra-compactness only occurs in the region of perturbative instability, for the model of bosonic stars we are analysing. Thus, it is not surprising that these models evolve into different configurations when perturbed. As we shall see in the evolutions presented here, ultra-compact bosonic stars decay into BHs, which is one of the possible fates already seen for unstable, but not ultra-compact, bosonic stars~\cite{Seidel:1990jh,Sanchis-Gual:2017bhw}. The simulations herein, moreover, allow us to: 1) establish this is a more generic fate in the ultra-compact case; 2) observe the timescale for the collapse, which turns out to be a short one. 
A different interesting question, which we shall not address herein since it is not the case for the solutions we are studying, is if there are any bosonic stars models, say, including rotation or self-interactions, for which ultra-compact bosonic stars are perturbatively stable.  

In the following we shall first consider the vector case and then, more briefly, the scalar case.
Our numerical evolutions employ a Cauchy approach. We introduce the $3+1$ decomposition of all dynamical quantities in the standard fashion (see, e.g.,~\cite{Gourgoulhon:2007ue,Alcubierre:2008,Cardoso:2014uka} for details). Concretely, we introduce the 3-metric
\begin{equation}
  \label{eq:3metric}
  \gamma_{\mu\nu} = g_{\mu \nu} + n_{\mu} n_{\nu} \ ,
\end{equation}
where $n^{\mu}$ denotes a timelike unit vector with normalization $n^{\mu} n_{\mu}=-1$.
The full spacetime metric $g_{\mu\nu}$ can then be expressed as
\begin{align}
\label{eq:LineElement}
ds^{2} & = g_{\mu\nu} dx^{\mu} dx^{\nu} \notag \\
       & = - \left( \alpha^{2} - \beta^{i} \beta_{i} \right) \, dt^{2}
             + 2 \beta_{i} \,dt \,dx^{i}
             +   \gamma_{ij} \, dx^{i} \,dx^{j}\,, 
\end{align}
where the lapse function $\alpha$ and shift vector $\beta^{i}$ describe the coordinate degrees of freedom.
%

\subsection{Proca stars} 
\label{sec41}

In the Proca case, the evolutions have been performed using the same code and setup employed in~\cite{Sanchis-Gual:2017bhw}, where we have also addressed spherical Proca star evolutions, albeit non-ultra-compact. The Einstein-Proca equations are formulated in the BBSN approach~\cite{Shibata:1995we,Baumgarte:1998te} and the usual choice of 1+log slicing condition and Gamma-driver shift conditions is made for the gauge equations. The code solves the evolution  equations in spherical symmetry using spherical polar coordinates and a second-order Partially Implicit Runge-Kutta (PIRK) method~\cite{Isabel:2012arx,Casas:2014}.

Following~\cite{Sanchis-Gual:2017bhw,Zilhao:2015tya}, the Proca field is split into its 3+1 variables, namely its scalar potential $\Phi$, 3-vector potentials $a_{i}$ and three-dimensional ``electric" ${\bf E}$ field. From the ansatz (\ref{ansatzfield}), 
the initial value for the Proca field variables is given as follows:
\begin{eqnarray}
\Phi&=&-n^{\mu}\mathcal{A}_{\mu}=-i\frac{V}{\alpha}\  , \label{propot}\\
a_{i}&=&\gamma^{\mu}_{i}\mathcal{A}_{\mu}=\frac{H_{1}}{r}\ ,\\
E^{i}&=&-i\,\frac{\gamma^{ij}}{\alpha}\,\biggl(D_{i} (\alpha\Phi)+\partial_{t}a_{j}\biggl)\nonumber\\
&=&i\frac{\gamma^{rr}}{\alpha}\,\biggl(D_{r} V+w\frac{H_{1}}{r}\biggl) \ .
\end{eqnarray}
Further details on the Einstein-Proca system can be found in~\cite{Sanchis-Gual:2017bhw,Zilhao:2015tya}.

For all evolutions presented herein, we have used the same logarithmic radial grid that extends from the origin to $r = 50$. We choose a time step of $\Delta t = C\Delta r$, where $C$ is the Courant factor.  For the ultra-compact models, the minimum resolution close to the origin is
$\Delta r = 0.0004$ and a Courant factor of $C=0.4$ for models PS1 and PS2, and $\Delta r = 0.0002$ and $C=0.8$ for PS3. It is worth pointing out the difficulty in performing long-term stable evolutions of these ultra-compact models as compared to the previous ones of~\cite{Sanchis-Gual:2017bhw}, due to the extreme accuracy requirements close to the origin. The minimum radial step and time step are two orders of magnitude smaller than in~\cite{Sanchis-Gual:2017bhw}, which involves a significant amount of computational time, even in spherical symmetry.

\begin{figure}[h!]
\centering
\includegraphics[height=2.6in]{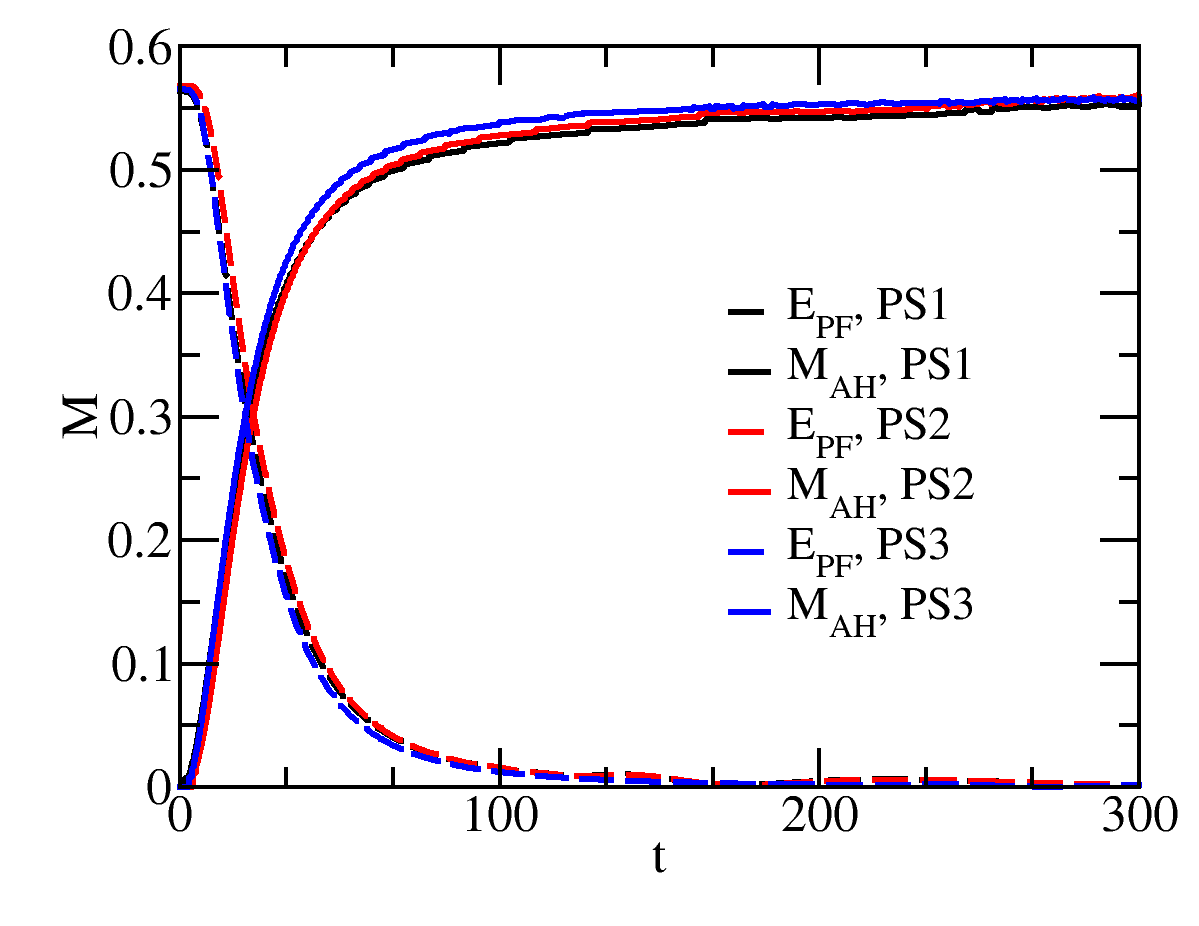} 
\includegraphics[height=2.6in]{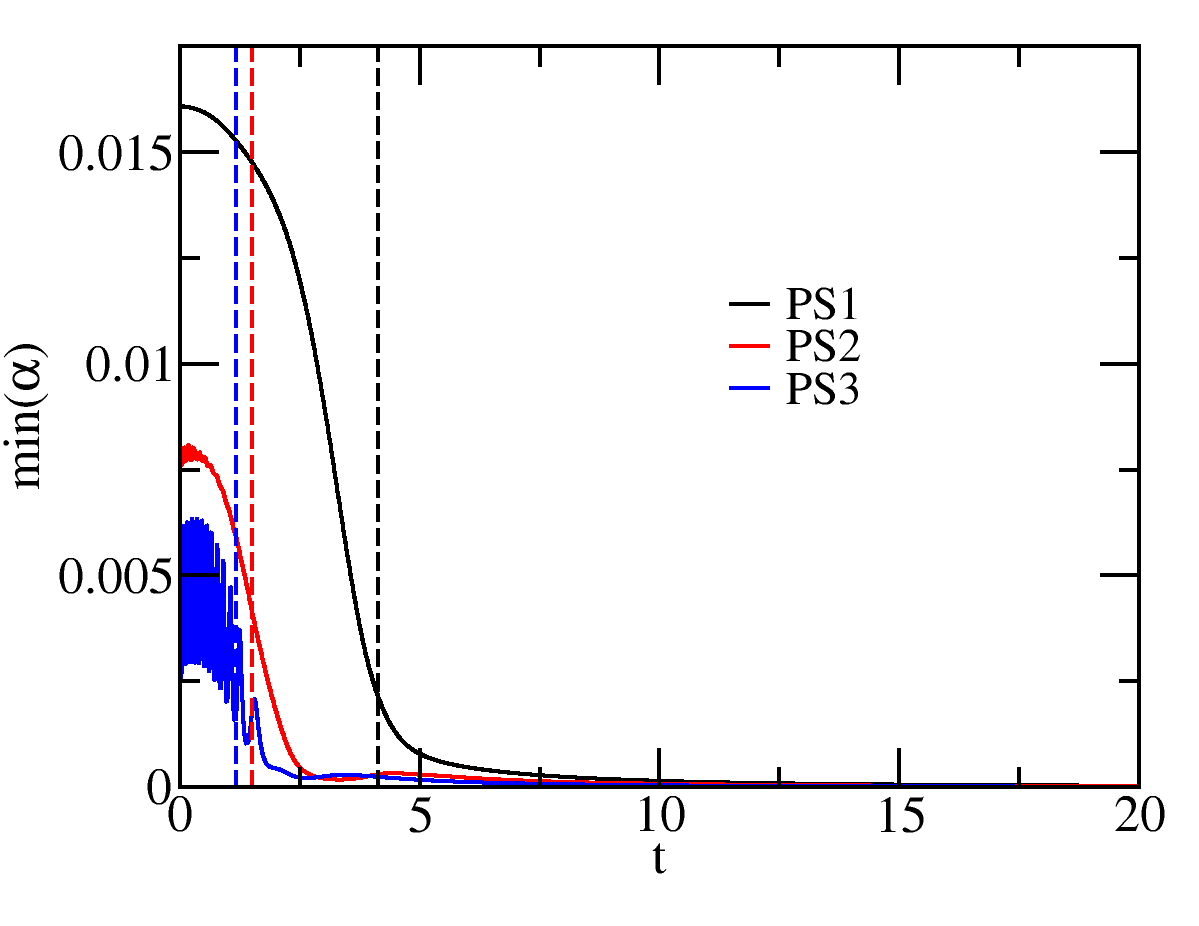}\\
\includegraphics[height=2.6in]{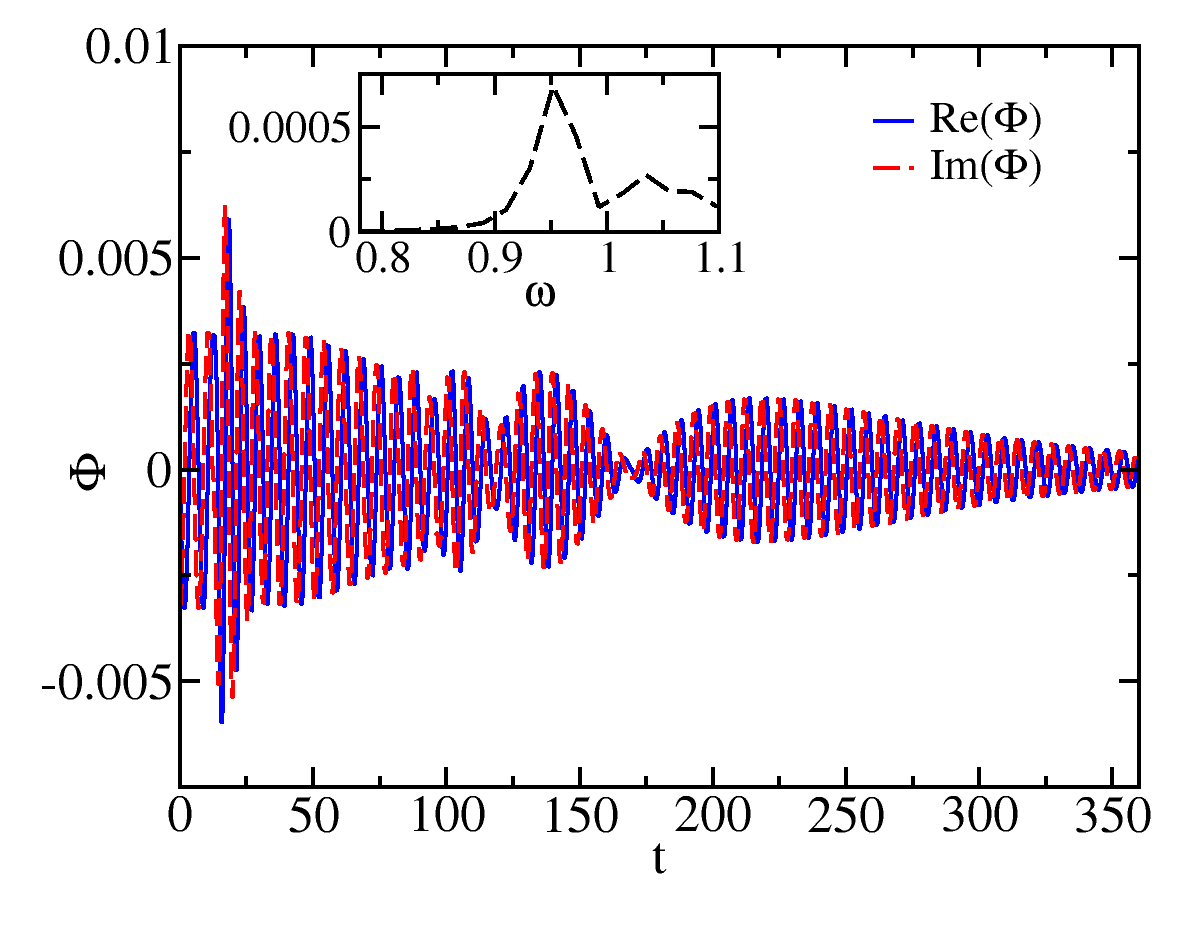}
\caption{(Top panel) Time evolution of the Proca field energy and AH mass for models 1, 2 and 3. (Middle panel) Time evolution of the minimal value of the lapse for all models.
  (Bottom panel, main) Time evolution of the amplitude of the scalar potential extracted at some point outside the horizon for model PS2. The inset shows the oscillation frequencies.}
\label{fig6}
\end{figure}

Figure~\ref{fig6} shows some of the main results concerning the evolutions of the models PS1-PS3. 
In the top panel of Fig.~\ref{fig6}  we exhibit the time evolution of the Proca field energy $E_{\rm PF}$ (see Eq.~(32) in~\cite{Sanchis-Gual:2017bhw}) and the irreducible mass of the apparent horizon (AH), once the latter forms in the evolutions.  The figure shows that for all three models, the discretization error of the numerical solution is sufficient to perturb the initial configurations and  trigger the collapse of the  ultra-compact Proca stars into BHs.   At  an early time in the evolution  an  AH forms.  This is confirmed in the middle panel, where the commencing time evolution of the minimum of the lapse function is shown. This so-called ``collapse-of-the-lapse"   is  a distinctive  feature of  AH  formation. The time coordinates for which those three AHs are found, computed using the AH finder implemented in the code of~\cite{Sanchis-Gual:2017bhw}, are indicated by the three vertical dashed lines in the middle panel of the figure. The corresponding times  are $(t_{\text{PS1}}=4.122$, $t_{\text{PS2}}=1.502$, $t_{\text{PS3}}=1.178)$ in units of $1/\mu$. As expected, the time scales of AH formation are smaller than those obtained for the much less compact models in~\cite{Sanchis-Gual:2017bhw}. The same trend we found in our previous work is also present in the ultra-compact case -- the more compact the model, the faster the AH appears.

Following AH formation, the Proca field energy evolves towards
being absorbed by the Schwarzschild BH, but within the timescale of our simulations some matter field energy remains outside the horizon.  This is confirmed in the bottom (main) panel of Fig.~\ref{fig6}, wherein we exhibit the time evolution of the amplitude of the scalar potential at some extraction radius outside the horizon and for model PS1.  The main plot shows that, when  the  BH  forms, a  part  of  the  Proca  field  remains  outside the horizon, with the real and imaginary parts oscillating with opposite phase. These are quasi-bound states of the Proca field around a Schwarzschild BH~\cite{Rosa:2011my,Zilhao:2015tya,Sanchis-Gual:2017bhw}.  Moreover, a beating pattern can be observed, a hint that more than one quasi-bound state of the Proca field is present outside the Schwarzschild BH. The corresponding frequencies of oscillation, obtained by Fourier-transforming the time evolution in the main panel, are displayed in the inset.

\begin{figure*}[tbhp]
\centering
\includegraphics[height=2.45in]{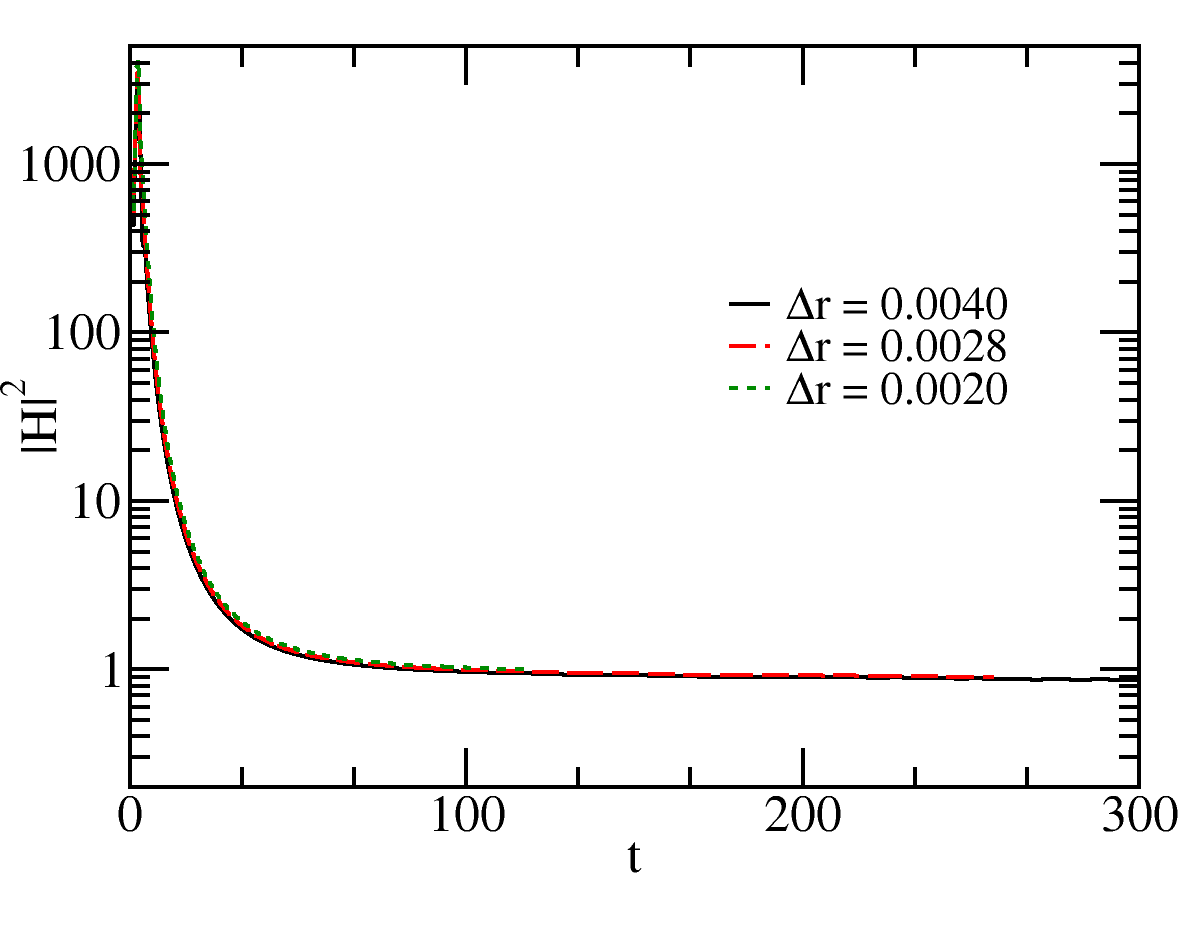}\includegraphics[height=2.45in]{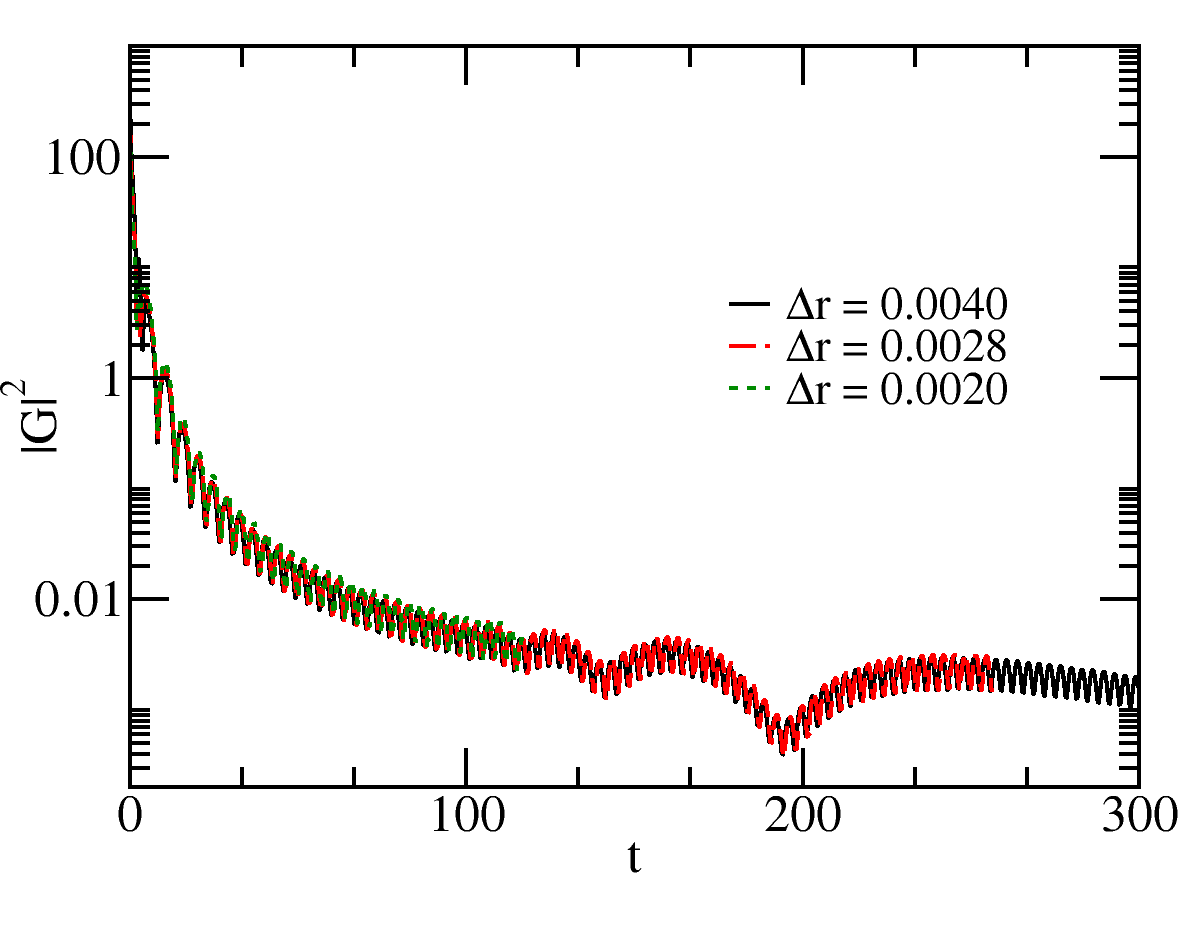}
\caption{Convergence analysis for model PS1 employing three different resolutions: $\Delta r=0.0004$, black curves, $\Delta r=0.0004/\sqrt{2}$, red curves, and $\Delta r=0.0004/2$, green curves. Left panel: L2 norm of the Hamiltonian constraint. Right panel: L2 norm of the Gauss constraint. The curves have been conveniently rescaled for second-order convergence, hence the overlapping.}
\label{fig7}
\end{figure*}

In order to assess the accuracy of our non-linear evolutions we have also performed a convergence test analysis for the Gauss constraint and the Hamiltonian constraint. The results are shown in Fig.~\ref{fig7} for model PS1. For both constraints we obtain the theoretical second-order convergence of the code.

To end this section, we comment that we have also evolved the ultra-compact Proca stars with an added perturbation, instead of relying solely on the discretization numerical error to trigger the non-linear dynamics. In~\cite{Sanchis-Gual:2017bhw} we observed this could change the fate of the star. For instance, by multiplying the Proca field by a number slightly smaller than one (namely $0.98$), therefore introducing a constraint-violating perturbation, we noticed that the fate of the star could change from collapsing into a BH to either migrating or dispersing (see Table III in~\cite{Sanchis-Gual:2017bhw}). In the present case of ultra-compact stars, however, even when multiplying the initial Proca field by $0.9$, the final fate of the non-linear dynamics is still gravitational collapse. This shows that ultra-compact bosonic configurations are much more prone to rapidly decay into a Schwarzschild BH, confirming our naive intuition. 

\subsection{Boson (scalar) stars} 
\label{sec42}

In the scalar case we use a completely independent numerical code and computational infrastructure. The code employed for the numerical evolutions makes use of the \textsc{EinsteinToolkit} infrastructure~\cite{Loffler:2011ay,EinsteinToolkit:web,Zilhao:2013hia}, which uses the \textsc{Cactus} Computational Toolkit~\cite{Cactuscode:web}, a software framework for high-performance computing. Mesh-refinement capabilities are handled by the 
\textsc{Carpet} package~\cite{Schnetter:2003rb,CarpetCode:web} and AHs are tracked with \textsc{AHFinderDirect}~\cite{Thornburg:2003sf,Thornburg:1995cp}.
The evolution of the spacetime metric is handled by \textsc{Lean}, originally presented in~\cite{Sperhake:2006cy} for vacuum spacetimes. Matter terms are here coupled straightforwardly using a separate thorn within the \textsc{Cactus} framework.

As in the code of~\cite{Sanchis-Gual:2017bhw} employed in the Proca case, \textsc{Lean} also uses the BSSN formulation of the Einstein
equations~\cite{Shibata:1995we,Baumgarte:1998te} with the moving puncture
method~\cite{Campanelli:2005dd,Baker:2005vv} and the usual 1+log slicing condition and Gamma-driver shift conditions for the gauge equations.
We employ the method-of-lines, where spatial derivatives are approximated by
fourth-order finite difference stencils, and we use the fourth-order Runge-Kutta scheme
for the time integration. Kreiss-Oliger dissipation is applied to evolved quantities in order to damp high-frequency noise.

Some details about the 3+1 decomposition of the Einstein-Klein-Gordon equations are presented in Appendix~\ref{appendixa}. Note that, whereas the code used for the evolutions of section~\ref{sec41} is a $1+1$ spherically-symmetric code, the one presented here evolves a 3-dimensional grid. This makes the numerical evolutions much more time-consuming. For these evolutions we have used an octant Cartesian grid extending from the origin to $x^i = 20$ ($x^i = x,y,z$, and we use units where we fix $\mu=1$) with 12 refinement levels. Resolution at the innermost refinement level was chosen to be $\Delta x^i = \frac{0.25}{2^{11}} \simeq 0.000122$ and time step $\Delta t = 0.4 \Delta x^i$. Such high resolution is needed to properly resolve the very steep gradients close to the origin due to the high compactness of the star. The spatial discretization is therefore two orders of magnitude smaller than the typical one used for instance in~\cite{Zilhao:2015tya}, making these evolutions extremely time-consuming.

\begin{figure}[h!]
\centering
\includegraphics[height=2.6in]{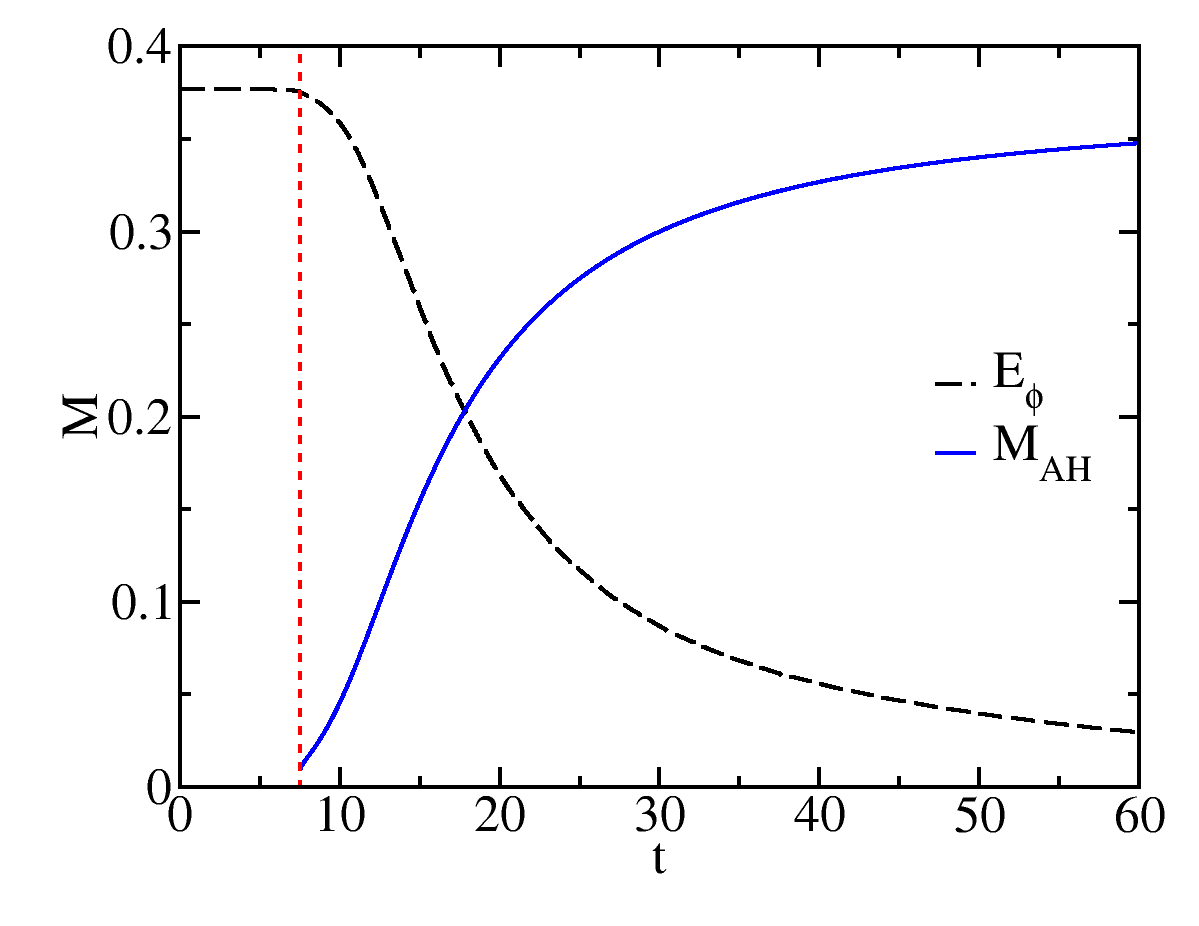} 
\includegraphics[height=2.6in]{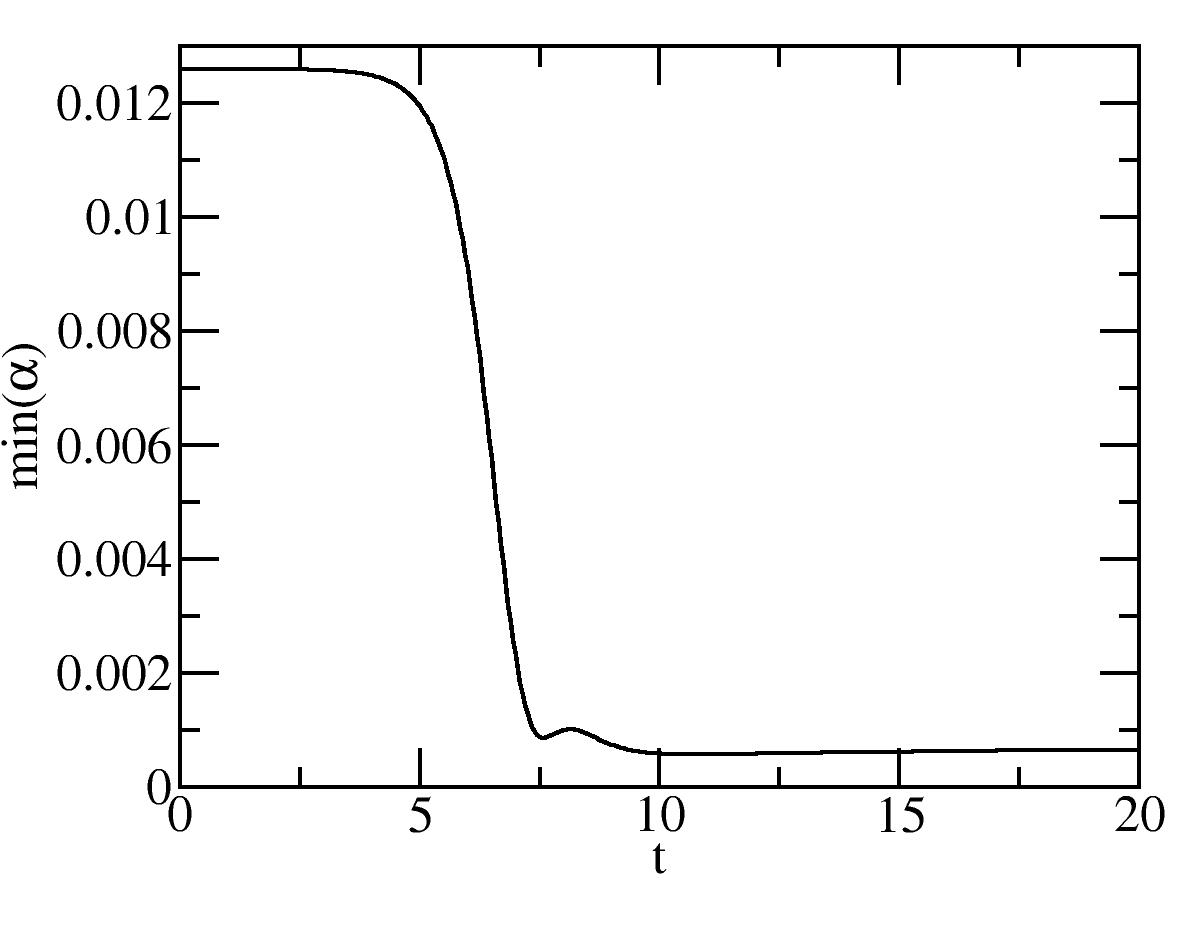}\\
\includegraphics[height=2.6in]{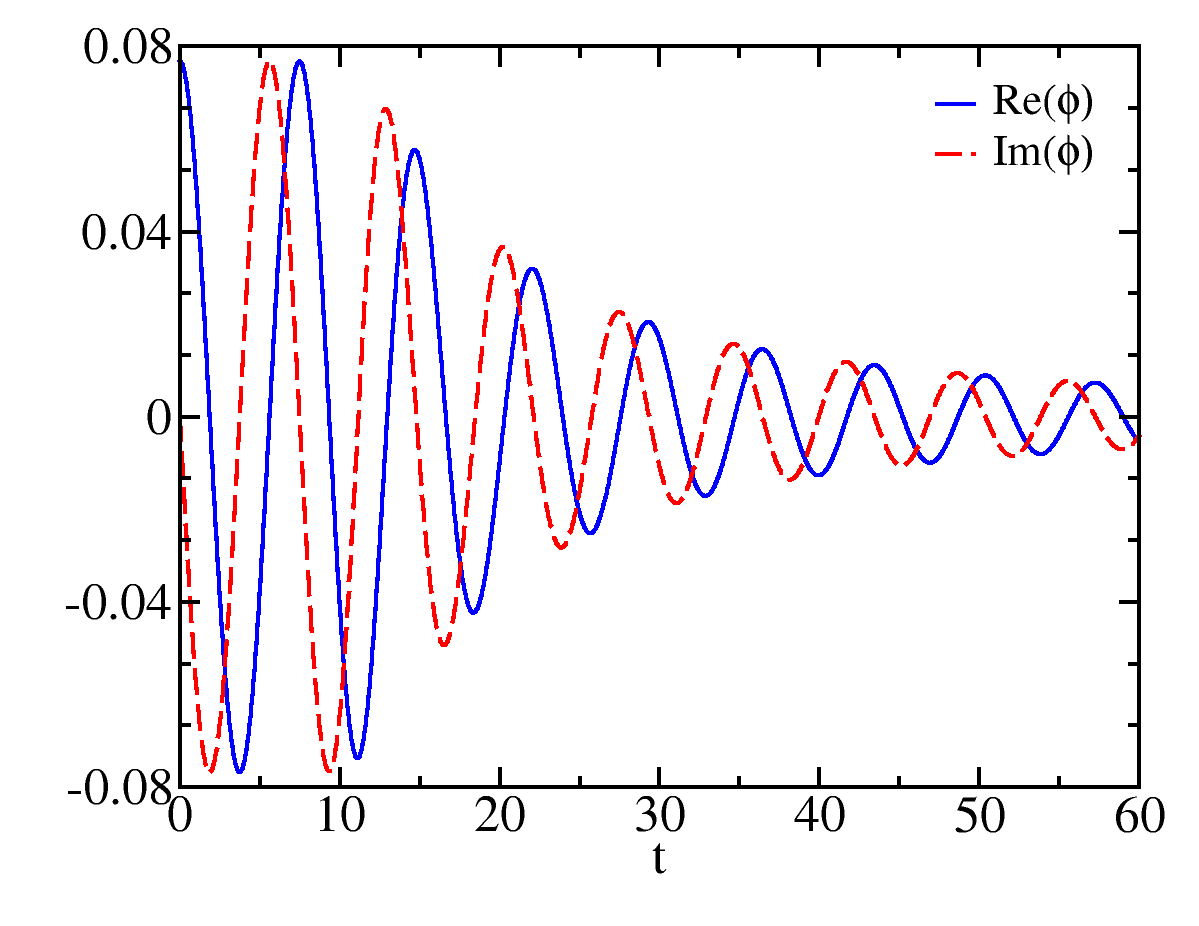}
\caption{(Top panel) Time evolution of the scalar-field energy and AH mass for model BS2. (Middle panel) Time evolution of the minimal value of the lapse for the same model.
  (Bottom panel) Time evolution of the real and imaginary part of the scalar field at a point outside the horizon.
}
\label{fig8}
\end{figure}

The results in the scalar case mimic closely the ones in the Proca case, despite the different spin  of the fundamental field that constitutes the star and the different code/computational infrastructure used in the solutions, showing the robustness of the conclusions. In Fig.~\ref{fig8} we exhibit analogous plots, in the scalar case, to those shown in Fig.~\ref{fig6} for the Proca case. In the top panel of Fig.~\ref{fig8} we show the time evolution of the scalar field
energy $E_{\Phi}$, computed as
\begin{equation}
  E_{\Phi} \equiv \int_{r>r_{\rm AH}} dx \, dy \, dz \, \alpha \, \sqrt{\gamma} \left( T_i^i - T_t^t \right) \,,
\label{eq:Ephi}
\end{equation}
and the irreducible
mass of the AH, once the latter forms in the evolution for model BS2.  Again, the
discretization error of the numerical solution is sufficient to trigger the collapse of
the  ultra-compact scalar star into a BH.   At  an early time in the evolution  an  AH
forms ($t_{\rm BS2}\approx 7.5$), which is marked by the red vertical dashed line in the plot. The formation of the horizon is again confirmed in the middle panel, where the time evolution of the minimum of the lapse function is shown. As for the Proca models, the minimum of the lapse function in the scalar case also  tends  to  zero,  signalling  AH  formation. Likewise, following AH formation, the scalar-field energy is increasingly absorbed by the Schwarzschild BH, but, as in the Proca case,  some of the energy remains outside the horizon within the timescale of our simulations.  This is confirmed in the bottom panel of Fig.~\ref{fig8}, wherein we display the time evolution of the scalar field at some extraction radius outside the horizon.   The real and imaginary parts oscillate with opposite phase corresponding to a quasi-bound start of the scalar field.

\begin{figure}[h!]
\centering
\includegraphics[height=2.6in]{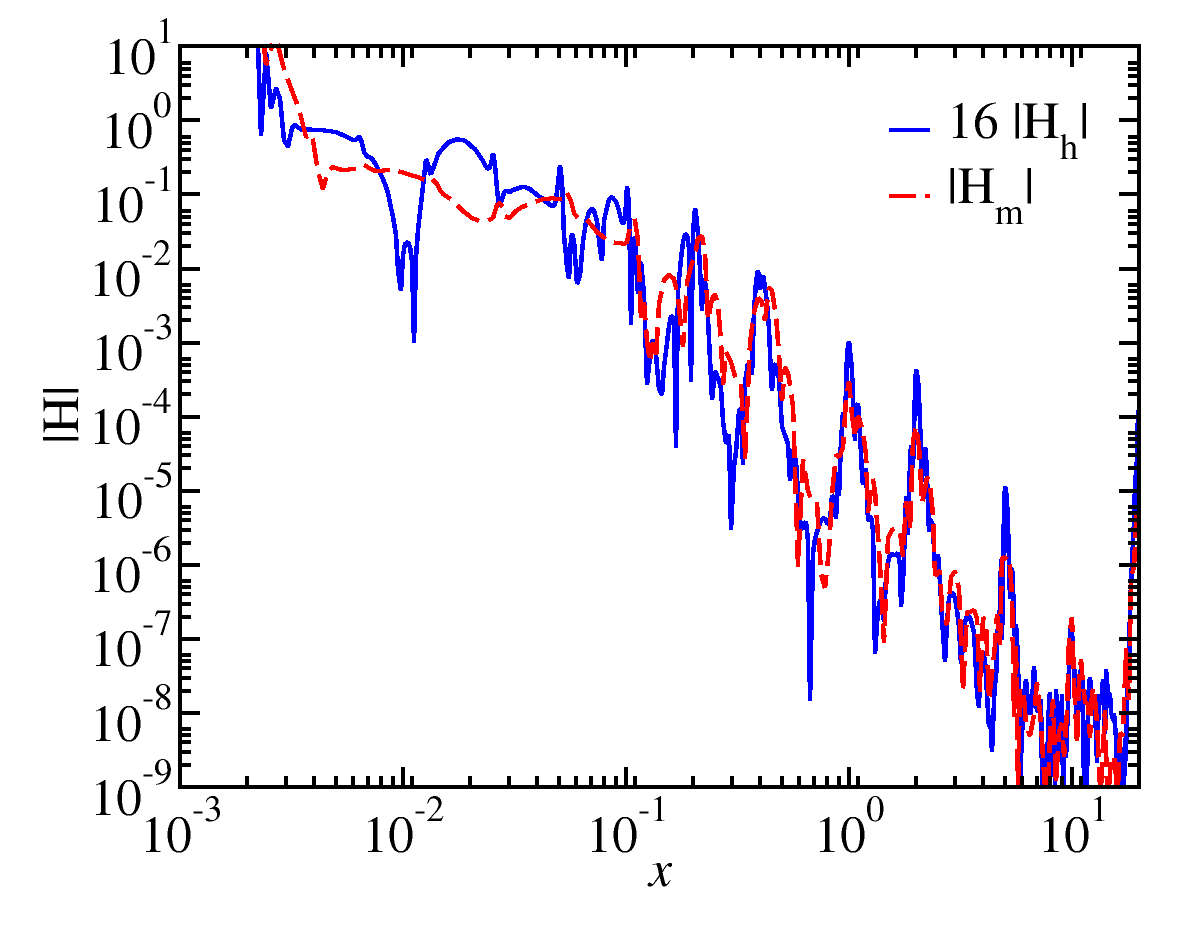}
\caption{Hamiltonian constraint violation at $t=10$ along the $x$-axis. The resolution at the innermost refinement level for the red curve is $\Delta x^i = 0.000244$, whereas for the blue curve $\Delta x^i = 0.000122$ was used. The blue curve has been multiplied by 16, the expected factor for fourth-order convergence. \label{fig:BS2_hc}}
\end{figure}

As we did in section~\ref{sec41} for the Proca case, we also assess the code employed in the 3D simulations of spherical (scalar) boson stars presented in this section. To this aim we show in
Fig.~\ref{fig:BS2_hc} the results of a convergence analysis on the Hamiltonian constraint violation at $t=10$ along the $x$-axis. As expected, we observe the theoretical fourth-order convergence of our code.

Let us conclude this section by remarking that, since these simulations are computationally very demanding, and take a very long time, we have only evolved models BS1 and BS3 up to horizon formation. We anticipate  their dynamics is identical to that of model BS2.

\section{Conclusions and final remarks}
\label{sec5}

The dawn of the era of gravitational-wave astronomy~\cite{Abbott:2016blz,Abbott:2016nmj,Abbott:2017vtc} promises to deliver detailed information about the nature of very compact objects in the Universe. The standard paradigm is that these are either BHs or neutron stars, but one cannot exclude, \textit{a priori}, the possibility that other compact objects, of an even more exotic nature, may hide in the Cosmos. 	

Intriguingly, it has been recently pointed out that the gravitational-wave ringdown of a horizonless UCO could be identical, at least in its initial stages, to that of a BH, since such ringdown stage is a signature of a LR, rather than of a horizon~\cite{Cardoso:2016rao}. Is therefore a risk of observationally mistaking UCOs by BHs and vice-versa, with current and near-future gravitational-wave measurements? 

In this paper we have tackled this question by considering a concrete UCO model. We have analysed some of the LR associated phenomenology of a class of exotic compact objects: scalar and vector boson stars. On the one hand, such compact objects are known to form from an incomplete gravitational collapse, and, at least in spherical symmetry, this formation is quite generic~\cite{Seidel:1993zk}. This contrasts with other UCO models proposed in the literature, for which no detailed formation scenario is known. On the other hand, within the family of solutions of these objects, some are ultra-compact, $i.e.$ they develop a LR. Since the spacetime geometry and matter field distribution for these bosonic stars are explicitly known they yield a privileged model to study the phenomenology of an UCO.

We have first analysed the lensing of light by the ultra-compact bosonic stars. Here, we have assumed that the scalar or vector matter (they are made of) interact very weakly with light, and thus the propagation of light is simply determined as the geodesic flow of the spacetime geometry. In other words, these bosonic stars are made of dark matter and thus they are permissive to electromagnetic radiation. Our analysis leads to two main conclusions, which are transversal to all explicit examples we have analysed. Firstly, the UCO produces no shadow, in contrast to a BH. Rather, it produces an annulus of darkness, associated to photon trajectories that come close to the LR and therefore take an arbitrarily long time to escape. This annular-like shadow, however, may be blurred into an apparent disk-like shadow, in an astrophysical context. Thus, this property, albeit a clear theoretical distinction from what happens for a BH, may not be a robust observable signature. Secondly, for comparable objects, the weak lensing region of the bosonic star and the BH is similar (measured, say, by the outermost Einstein ring), but the strong lensing region of the star is considerably smaller than that of a BH, under similar observational conditions. Thus, an ultra-compact bosonic star and a Schwarzschild BH with the same mass, observed at a similar distance, will be distinguishable. That is, even if the lensing of the bosonic star produces an effective disk-like shadow, due to the blurring of the annulus-like strong lensing region, this disk is considerably smaller than that of the shadow of the comparable Schwarzschild BH ($e.g$ $\sim$ 6 times smaller for model PS2). Thus, it seems that even if the existence of a LR for a horizonless object can mimic a part of its gravitational wave relaxation signal, it does not mimic (simultaneously), its electromagnetic phenomenology.

After analysing the lensing of light we have considered the dynamics of a perturbed ultra-compact bosonic star. It was clear from the outset that, when in the ultra-compact regime, the bosonic stars of the model we are considering (sometimes called mini-boson stars in the scalar case~\cite{Schunck:2003kk}) are unstable. Our fully non-linear numerical simulations have allowed us to establish two conclusions about their fate. Firstly, they collapse into a BH and this is a quite robust fate. That is, even varying the perturbation, which for less compact stars could change their fate into a migration or dispersion~\cite{Sanchis-Gual:2017bhw}, ultra-compact stars still collapse into a BH. Secondly, the collapse is fast, occurring within a few light crossing times.

Overall, the two aspects analysed in this paper (lensing and dynamics) of this model of UCOs, emphasises that it is quite challenging for a UCO model to mimic all the phenomenology and dynamics of a BH, at least in spherical symmetry, even if it can mimic some of it (like its ringdown). It would be quite interesting to see if by changing the model of bosonic stars, $e.g.$ by introducing self-interactions or rotation,  the conclusions we have obtained herein for the spherically symmetric mini-bosonic stars change.

Finally, we would like to emphasise that the UCOs we have considered actually possess two LRs. The phenomenology described herein is associated with the outermost unstable one. But the existence of an innermost stable one also may have important dynamical consequences, in particular with respect to the spacetime stability~\cite{Keir:2014oka,Cardoso:2014sna}. It has been recently proven~\cite{Cunha:2017qtt} that for generic horizonless UCOs forming smoothly from incomplete gravitational collapse, within physically reasonable models, this stable LR is always present, and therefore the instability it may trigger, again challenging the possibility of physically realistic horizonless UCOs in the Universe.

\section*{Acknowledgements}

P.C. is supported by Grant No. PD/BD/114071/2015 under the FCT-IDPASC Portugal Ph.D. program.
M.Z. acknowledges support by the ERC Starting Grant HoloLHC-306605 and FCT (Portugal) IF programme IF/00729/2015.
This work has been supported by the Spanish MINECO (grant AYA2015-66899-C2-1-P), 
by the Generalitat Valenciana (PROMETEOII-2014-069, ACIF/2015/216), by the FCT (Portugal) IF programme, by the CIDMA (FCT) 
strategic project UID/MAT/04106/2013 and by  the  European  Union's  Horizon  2020  research  and  innovation  programme  under  the  Marie
Sklodowska-Curie grant agreement No 690904 and by the CIDMA project UID/MAT/04106/2013. Computations have been 
performed at the Servei d'Inform\`atica de la Universitat de Val\`encia, at the Blafis cluster at the University of Aveiro and at the Baltasar cluster at IST.

\appendix

\section{3+1 decomposition for the Einstein-Klein-Gordon system} 
\label{appendixa}

Let us briefly present the $3+1$ decomposed equations of motion in the Einstein-Klein-Gordon case. To complete the characterization of the full spacetime we define the extrinsic curvature
\begin{equation}
\label{eq:KijDef}
K_{ij}  =   - \frac{1}{2\alpha} \left( \partial_{t} - \Lie_{\beta} \right) \gamma_{ij} \ ,
\end{equation}
and analogously introduce the ``canonical momentum'' of the complex scalar field $\Phi$
\begin{equation}
\label{eq:Kphi}
K_{\Phi} = -\frac{1}{2\alpha}  \left( \partial_{t} - \Lie_{\beta} \right) \Phi \,,
\end{equation}
where $\Lie$ denotes the Lie derivative.
Our evolution system can then be written in the form
%
%
\begin{subequations}
\begin{align}
  \p_{t} \gamma_{ij} & = - 2 \alpha K_{ij} + \Lie_{\beta} \gamma_{ij} \,,
                       \label{eq:dtgamma} \\
  \p_{t} K_{ij}      & =  - D_{i} \p_{j} \alpha
                       + \alpha \left( R_{ij} - 2 K_{ik} K^{k}{}_{j} + K K_{ij} \right) \nonumber \\
                       & \quad + \Lie_{\beta} K_{ij} 
                       + 4\pi \alpha \left[ (S-\rho) \gamma_{ij} - 2 S_{ij} \right] \,,
                                         \label{eq:dtKij} \\
  \p_{t} \Phi & = - 2 \alpha K_\Phi + \Lie_{\beta} \Phi\
                \,, \label{eq:dtPhi} \\
  \p_{t} K_\Phi &  = \alpha \left( K K_{\Phi} - \frac{1}{2} \gamma^{ij} D_i \partial_j \Phi
                  + \frac{1}{2} \mu^2 \Phi \right) \nonumber \\
                 & \quad - \frac{1}{2} \gamma^{ij} \partial_i \alpha \partial_j \Phi
                       + \Lie_{\beta} K_\Phi \,, \label{eq:dtKphi}
\end{align}
\end{subequations}
%
where $D_i$ is the covariant derivative with respect to the $3$-metric.

For the numerical simulations we re-write the evolution equations above in the BSSN scheme~\cite{Shibata:1995we,Baumgarte:1998te}, which renders the system well-posed.
The full system of evolution equations is then
\begin{subequations}
  \allowdisplaybreaks
  \label{eq:bssn}
  \begin{align}
    \partial_t \tilde{\gamma}_{ij} & = \beta^k \partial_k \tilde{\gamma}_{ij} +
    2\tilde{\gamma}_{k(i} \partial_{j)} \beta^k - \frac{2}{3}
    \tilde{\gamma}_{ij} \partial_k \beta^k -2\alpha \tilde{A}_{ij},  \\
    \partial_t \chi & = \beta^k \partial_k \chi + \frac{2}{3} \chi (\alpha K
    - \partial_k \beta^k),  \\
    \partial_t \tilde{A}_{ij} & = \beta^k \partial_k \tilde{A}_{ij}
      + 2\tilde{A}_{k(i} \partial_{j)} \beta^k
      - \frac{2}{3} \tilde{A}_{ij} \partial_k \beta^k  \notag \\
      & \quad + \chi \left( \alpha R_{ij} - D_i \partial_j \alpha\right)^{\rm
      TF}
      + \alpha \left( K\,\tilde{A}_{ij}
      - 2 \tilde{A}_i{}^k \tilde{A}_{kj} \right) \notag \\
      & \quad - 8 \pi \alpha \left(
          \chi S_{ij} - \frac{S}{3} \tilde \gamma_{ij}
        \right), \\
    \partial_t K & = \beta^k \partial_k K - D^k \partial_k \alpha + \alpha \left(
                   \tilde{A}^{ij} \tilde{A}_{ij} + \frac{1}{3} K^2 \right) \notag \\
    & \quad  + 4 \pi \alpha (\rho + S), \\
    \partial_t \tilde{\Gamma}^i & = \beta^k \partial_k \tilde{\Gamma}^i
       - \tilde{\Gamma}^k \partial_k \beta^i + \frac{2}{3}
    \tilde{\Gamma}^i \partial_k \beta^k + 2 \alpha \tilde{\Gamma}^i_{jk}
                                  \tilde{A}^{jk} \notag \\
                                  & \quad + \frac{1}{3} \tilde{\gamma}^{ij}\partial_j \partial_k
    \beta^k
    + \tilde{\gamma}^{jk} \partial_j \partial_k \beta^i \nonumber \\
    & \quad - \frac{4}{3} \alpha \tilde{\gamma}^{ij} \partial_j K -
    \tilde{A}^{ij} \left( 3 \alpha \chi^{-1} \partial_j \chi + 2\partial_j
      \alpha \right) 
                    \notag \\
    & \quad - 16 \pi \alpha \chi^{-1} j^i \label{eq:tilde-Gamma-evol} \,, \\
  \p_{t} \Phi & = - 2 \alpha K_\Phi + \Lie_{\beta} \Phi\
                \,, \label{eq:dtPhi-BSSN} \\
  \p_{t} K_\Phi &  = \alpha \left( K K_{\Phi} - \frac{1}{2} \gamma^{ij} D_i \partial_j \Phi
                  + \frac{1}{2} \mu^2 \Phi \right) \nonumber \\
                 & \quad - \frac{1}{2} \gamma^{ij} \partial_i \alpha \partial_j \Phi
                       + \Lie_{\beta} K_\Phi \,, \label{eq:dtKphi-BSSN}
  \end{align}
\end{subequations}
%
%
with the source terms given by
\begin{equation}
  \label{eq:source}
   \begin{aligned}
  \rho & \equiv T^{\mu \nu}n_{\mu}n_{\nu} \,,\\
  j_i  &\equiv -\gamma_{i\mu} T^{\mu \nu}n_{\nu} \,, \\
  S_{ij} &\equiv \gamma^{\mu}{}_i \gamma^{\nu}{}_j T_{\mu \nu} \,, \\
  S     & \equiv \gamma^{ij}S_{ij} \,.
   \end{aligned}
\end{equation}
For reasons of convenience, we evolve the real and imaginary part of the scalar field $\Phi$ as separate, independent, variables.
Finally, the evolution is subject to a set of constraints given by
\begin{align}
\label{eq:Hamiltonian}
\mathcal{H} & \equiv R - K_{ij} K^{ij} + K^2 - 16 \pi \rho
       = 0\,,\\
\label{eq:momentumConstraint}
\mathcal{M}_{i} & \equiv D^{j} K_{ij} - D_{i} K 
        - 8\pi j_i
       = 0 \,.
\end{align}

\bibliography{num-rel}

\end{document}